\documentclass[10pt]{revtex4}

\usepackage[english]{babel}
\usepackage{etex}
\usepackage{graphicx}
\usepackage{float}
\usepackage{chemarr}
\usepackage{times}
\usepackage{systeme}
\usepackage[version=3]{mhchem} 
\usepackage[utf8]{inputenc}
\usepackage[a4paper,margin=2.cm,footskip=.5cm]{geometry}
\usepackage{hyperref}

\usepackage{amsmath}
\usepackage{amssymb}
\usepackage[thinlines]{easybmat}
\usepackage[warning,math]{easyeqn}

\usepackage{times,mathptmx}

\usepackage{balance} 

\usepackage{lastpage}

\newcommand{\ds}{\displaystyle}

\newcommand{\de}{\partial}
\newcommand{\1}{\mathbb{I}\hspace*{-0.4ex}}
\newcommand{\bmr}{\mbox{\boldmath$r$}}

\newcommand{\LB}{\mbox{\boldmath$L$}}

\newcommand {\Be}{\begin{eqnarray*}}
\newcommand {\Ee} {\end{eqnarray*}}
\newcommand {\bey} {\begin{eqnarray}}
\newcommand {\eey} {\end{eqnarray}}
\newcommand{\bit}{\begin{itemize}}      
\newcommand{\eit}{\end{itemize}}
\newcommand{\bfl}{\begin{flusleft}}
\newcommand{\efl}{\end{flusleft}}
\newcommand{\bfr}{\begin{flushright}}
\newcommand{\bc}{\begin{center}}
\newcommand{\ec}{\end{center}}
\newcommand{\ben}{\begin{enumerate}}    
\newcommand{\een}{\end{enumerate}}

\def\a{\alpha}

\def\b{\beta}

\def\ve{\varepsilon}

\def\<{\langle}
\def\>{\rangle}

\newcommand{\be}{\begin{equation}}
\newcommand{\ee}{\end{equation}}

\begin{document}

\title{Reaction rate of a composite core-shell nanoreactor 
       with multiple, spatially distributed embedded nano-catalysts}                   
\vspace{0.6cm}

%

\author{Marta Galanti}
\affiliation{Universit\'e d'Orl\'eans, Ch\^ateau de la Source, 45100, Orl\'eans, France,\\
                             Centre de Biophysique Mol\'eculaire, CNRS-UPR4301,
                             Rue C. Sadron, 45071, Orl\'eans, France.}
\affiliation{Universit\`a degli Studi di Firenze, Dipartimento di Fisica e Astronomia and CSDC, 
                             via G. Sansone 1, IT-50019 Sesto Fiorentino, Firenze, Italia.}
\author{Duccio Fanelli}
\affiliation{Universit\`a degli Studi di Firenze, Dipartimento di Fisica e Astronomia and CSDC, 
                             via G. Sansone 1, IT-50019 Sesto Fiorentino, Firenze, Italia.}
\author{Stefano Angioletti-Uberti}                       
\affiliation{International Research Centre for Soft Matter, \\Beijing University of Chemical Technology, 
                             Beijing 100029, China.}
\author{Matthias Ballauff} 
\affiliation{Institut f\"ur Physik, Humboldt Universit\"at zu Berlin, 12489 Berlin, Germany.}
\affiliation{Institut f\"ur Weiche Materie und Funkionale Materialien, \\
                             Helmholtz-Zentrum Berlin, 14109 Berlin, Germany.}
\author{Joachim Dzubiella} 
\affiliation{Institut f\"ur Physik, Humboldt Universit\"at zu Berlin, 12489 Berlin, Germany.}
\affiliation{Institut f\"ur Weiche Materie und Funkionale Materialien, \\
                             Helmholtz-Zentrum Berlin, 14109 Berlin, Germany.}    
\author{Francesco Piazza}
\affiliation{Universit\'e d'Orl\'eans, Ch\^ateau de la Source, 45100, Orl\'eans, France,\\
                             Centre de Biophysique Mol\'eculaire, CNRS-UPR4301,
                             Rue C. Sadron, 45071, Orl\'eans, France.}                                                       


\begin{abstract}
We present a detailed theory for the total reaction rate constant of a composite core-shell
nanoreactor, consisting of a central solid core surrounded by a hydrogel layer of variable thickness, 
where  a given number of small catalytic nanoparticles are embedded at prescribed positions and 
are endowed with a prescribed surface reaction rate constant.  Besides the precise geometry of 
the assembly, our theory accounts explicitly for the diffusion coefficients of the reactants 
in the hydrogel and in the bulk as well as for their transfer free energy 
jump upon entering the hydrogel shell. Moreover, we work out an approximate analytical 
formula for the overall rate constant, which is valid in the physically relevant range 
of geometrical and chemical parameters. We discuss in depth how the diffusion-controlled part of the rate
depends on the essential variables, including the size of the central core. 
In particular, we derive some simple rules for estimating the 
number of nanocatalysts per nanoreactor for an efficient catalytic performance in the 
case of small to intermediate core sizes.  
Our theoretical treatment promises to provide a very useful and flexible tool for the design of superior 
performing  nanoreactor geometries and with optimized nanoparticle load. 
\end{abstract}

\maketitle

\section{Introduction}

In recent years, metallic nanoparticles have emerged as potent catalysts for various 
applications~\cite{Pushkarev2012,Zhang2012}. In particular, the discovery that gold becomes 
a catalyst when divided to the nanophase has led to an intense research in this 
field~\cite{Haruta2003,Hutchings2005}. In many cases the synthesis and the catalytic applications 
must be handled in the liquid phase, mostly in water. Secure handling of nanoparticles in a 
liquid phase can be achieved by polymeric carriers that have typical dimensions in the colloidal domain. 
Examples thereof include dendrimers~\cite{Crooks2001,Noh2015} or spherical polyelectrolyte 
brushes~\cite{Ballauff2007}. Such systems allow one to generate nanoparticles in aqueous phase 
in a well-defined manner and handle them securely in catalytic reactions. \\
\indent More recently, thermosensitive colloidal microgels have been used as carriers for 
metallic nanoparticles in catalysis~\cite{Lu2011}. Fig.~\ref{f:scheme} displays the scheme 
of such a carrier system that may be regarded as a nanoreactor: a thermosensitive network 
composed of cross-linked chains of poly(N-isopropylacrylamide) (PNIPAM) has been attached to a solid 
core made of an inert material as, {\em e.g.}, polystyrene or silica~\cite{Lu2006c}. 
Metal nanoparticles are embedded in the network which is fully swollen in cold water. 
Raising the temperature above the critical temperature (32$^\circ$ C for PNIPAM), a volume transition 
takes place within the network and most of the water is expelled~\cite{Lu2011}. 
Lu {\it et al.}~\cite{Lu2006c} have been the first to show that the catalytic activity of 
the embedded nanoparticles is decreased when shrinking the network by raising the temperature. 
This effect has been explained by an increased diffusional resistance mass transport 
within the shrunk network~\cite{Lu2006c,Lu2011}. A similar model has been advanced by 
Carregal-Romero {\it et al.} when considering the catalytic activity of a single 
gold nanoparticle embedded concentrically in a PNIPAM-network~\cite{marzan1}.\\
\indent Recently, we have shown that the mobility of reactants is not the only important factor:
an even larger role is played by the change of polarity of the network when considering 
mass transport from bulk to the catalyst(s) through such medium~\cite{Wu2012,Angioletti-Uberti2015}. 
This theory is based on the well-known seminal paper by Debye~\cite{debye42} and considers 
a single nanoparticle located in the center of a hollow thermosensitive network~\cite{Angioletti-Uberti2015}. 
Here, the substrate that reacts at the surface of the nanoparticle diffuses through a 
free-energy landscape created by the hydrogel environment. In other words, 
the reactants experience a change in the solvation free energy when entering the gels from bulk solvent, 
which can be equally regarded as adsorption free energy or {\em transfer} free energy.  
For instance, the free energy of a substrate may
be lowered upon entering the network. In this way the number of substrate molecules  
in the network will be augmented, so that their increased concentration in the vicinity of the catalyst 
will lead to a higher reaction rate. The free-energy change $\Delta G \stackrel{\rm def}{=} G_{\rm in} - G_{\rm out}$ 
for the substrate outside and inside the network leads to a Nernst distribution for the substrate's 
concentration within the system.  This effect offers a new way to manipulate the catalytic activity and 
selectivity of metallic nanoparticles~\cite{Wu2012}.\\
\indent In this paper we formulate a more general theory, that is able to account for 
the geometry of core-shell nanoreactors featuring {\em many} catalysts,
as shown schematically in Fig.~\ref{f:scheme}.  Here, a given number $N$ of catalytic centers 
are encapsulated randomly in a network.  We calculate the total rate of the catalytic reaction 
for a prescribed geometry of the catalysts, given values of $N$ and $\Delta G$ and specified 
diffusion constants for the substrate in the bulk and in the network. The rate constant computed in this way 
can be compared to that characterizing an equal number of particles  suspended freely in solution.
The present model has been designed to describe the well-studied core-shell systems~\cite{Lu2006c, Lu2011}, 
but is equally adapted to the study of systems where catalytic centers are embedded in homogeneous microgels~\cite{Shi2014}. 
Up to now, most of the experimental work has been done using the reduction of 4-nitrophenol by borohydride ions 
in aqueous solution~\cite{Aditya2015,Zhao2015}. This reaction can be regarded as a model reaction~\cite{Herves2012},
since it can be monitored with high precision thus leading to very accurate kinetic data. 
The rate-determining step proceeds at the surface of the nanoparticles and the 
mechanism is known~\cite{Gu2014}. The present theory, however, comprises also the nanoreactors in which enzymes 
are used as catalytic centers embedded in the network shell~\cite{Welsch2012}. \\
\indent In general, diffusion-influenced reactions (DIR) are ubiquitous in many contexts in physics, chemistry and 
biology~\cite{Calef:1983aa,Rice:1985kx,Szabo:1989fk,Zhou:2008vf}.
However, while the mathematical foundations for the description of DIR in simple systems have been laid 
nearly a century ago~\cite{Smoluchowski:1916fk,debye42,Collins:1949aa}, many important present-day problems,
including the catalytic activity of composite core-shell nanoreactors, 
require considering complex geometries and multi-connected reactive boundary systems. 
The first attempts to consider DIR featuring many competing sinks date back to the 
1970s~\cite{Deutch:1976aa,Felderhof:1976aa}, while more sophisticated methods have been 
developed subsequently to deal with arbitrary systems comprising many partially reactive 
boundaries~\cite{Traytak:1992,Traytak:2003aa,Gordeliy:2009aa}.  
Along similar lines, the theory developed in this paper, 
based on general results proved in Ref.~\cite{Traytak:2003aa}, 
provides a novel, accurate description of DIR occurring between a small substrate 
molecule and the catalytic centers embedded in a large, composite nanoreactor system. \\
%
%
\indent Our theory is fully general, in that it covers the whole spectrum of rate-limiting 
steps in catalysis, from reaction-limited to diffusion-limited reactions. While the theory allows
one to compute the reaction rate for an arbitrary catalytic surface turnover rate, closed-form 
analytical expression are derived for strongly reaction-limited and diffusion-limited reactions.
In the limit of a dilute random distribution of NPs encapsulated in a thick hydrogel shell, 
we find that the overall diffusion-controlled  rate constant of our core-shell  composites
is described by a Langmuir-like isotherm of the form  
\begin{equation}
\label{e:Langm}
\frac{k}{k_S} = 
                  \frac{ N\ve\,\zeta e^{-\beta \Delta G}}
                       {  1 + N\ve\,\zeta e^{-\beta \Delta G}}
\end{equation} 
where $N$ is the number of nanoparticles, $\zeta = D_i/D_o$ is the ratio of the diffusion constants 
in the hydrogel ($i$ for inner) and bulk ($o$ for outer), $\ve = a/R_0 \ll 1$ is the ratio of 
NP size (radius) to the nanoreactor size and  $k_S = 4\pi D_o R_0$ is the Smoluchowski rate constant for 
the nanoreactor as a whole, {\em i.e.}
the total flux (in units of bulk substrate concentration) of substrate molecules to a stationary 
perfectly  absorbing sink of size $R_0$ in the bulk. The above expression is valid for 
small sizes of the central core. Interestingly, for configurations where the core size becomes of the same 
order of the whole composite (thin shell), our theory shows that in general the rate constant is
increased, up to 40 \%, depending on the transfer free energy jump and on the 
reactant mobility in the shell.\\
\indent In the limit of slow surface substrate-product conversion rate, i.e. for reaction-limited 
kinetics, we find that  
\begin{equation}
\label{e:ratesmallksant}
k \simeq Nk^\ast e^{-\beta \Delta G} + \mathcal{O}[(k^\ast/k_S^+)^2]
\end{equation}
where $k^\ast$ is the intrinsic turnover rate constant that describes the 
transformation of substrate to product molecules at the nanocatalyst surface 
(units of inverse concentration times inverse time). 
This means that when the surface substrate conversion rate constant is weak, 
the geometrical features of the overall assembly and  the mobility  
of substrate molecules within the hydrogel shell become immaterial. 
In this case, the crucial control parameter is the transfer free energy jump.
\indent The paper is organized as follows. In section~\ref{sec:model} we describe our mathematical 
model and pose the associated boundary-value problem. In section~\ref{sec:sol}, we describe concisely
the procedure that leads us to the exact solution of the posed problem (the mathematical details 
can be found in the appendix). Section~\ref{sec:analytic}
illustrates an analytic approximation that 
provides an extremely good description 
of the exact solution for small core sizes in the physically relevant range of parameters.
In particular, we discuss how this formula can be used to derive practical criteria 
to design nanoreactors with optimized performances.
Finally, we wrap up our main results in section~\ref{sec:summary}. 

%
%
%
%
%
\section{Core-shell model and defining equations\label{sec:model}}
%
\begin{figure}[t!]
\centering
\includegraphics[width=10truecm]{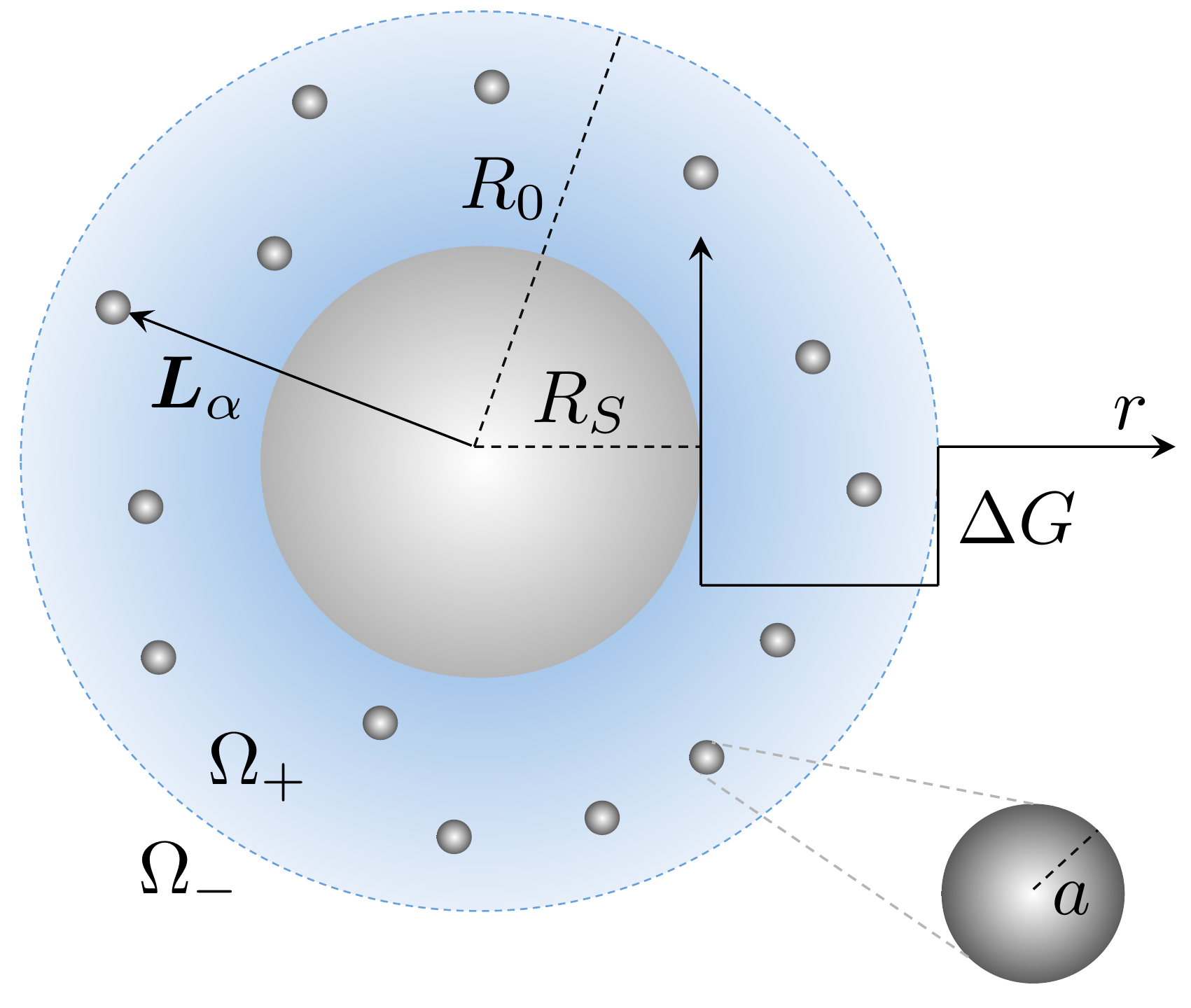}
\caption{Scheme of a core-shell nanoreactor of radius $R_{0}$ 
containing $N+1$ spheres:
the solid polystyrene (PS) core (radius $R_{S}$) is shown at the 
center, along with  $N$ catalytic nanoparticles of radius $a$ at 
positions $\mathbf{L}_\a$ ($\alpha=2,3,\dots,N+1$).
The internal (microgel) domain $\Omega_+$ (with reactant diffusion coefficients $D_{i}$) and 
the external (bulk solution) domain $\Omega_-$ (with reactant diffusion coefficients $D_{o}$) 
are indicated explicitly, together with a schematic free-energy radial profile
showing the transfer free-energy jump $\Delta G$. 
In our treatment the latter can be both repulsive, $\Delta G>0$, or attractive, $\Delta G<0$. 
\label{f:scheme}
}
\end{figure}
%
We model a core-shell nanoreactor consisting of a polystyrene (PS) core surrounded by a microgel 
layer as two concentric spheres centered at the origin of a Cartesian 3$D$ frame,
as depicted in  Fig.~\ref{f:scheme}. We denote with $R_{S}$ and $R_{0}$ the 
core and shell radius, respectively. The shell is assumed to be a  homogeneous continuum, 
carrying $N$ small nano-catalysts (metal nanoparticles or enzymes) that we model as spheres of radius $a$. 
For the sake  of simplicity, and in accordance to our general multi-sink theory~\cite{Galanti:2016ab},
we label the PS core as the inner sphere with $\a=1$ and position vector $\LB_{1}=0$ and
denote the position of the  $N$ nanocatalysts with the   vectors $\LB_{\a}$, $\a=2,3,\dots,N+1$. We want to compute the 
total reaction rate constant for reactions where a substrate (or ligand) molecule is converted 
to some product species at the surface of the catalyst spheres. 
These are endowed with a surface rate constant $k^\ast$, which is in general a function of temperature due to underlying 
thermally activated surfaces processes. Let us  denote with $\mathcal{S}_{0}\equiv\{r_{0},\theta_{0},\varphi_{0}\}$ 
the reference frame with the origin 
at the nanoreactor center and with $\mathcal{S}_{\a}\equiv \{r_{\a},\theta_{\a},\varphi_{\a} \}$
the $N$ reference frames with the origins at the nanospheres centers and the axes parallel to 
$\mathcal{S}_{0}$ (of course $\mathcal{S}_{1}\equiv\mathcal{S}_{0}$). 
This formally defines the  following $3D$ domains
\begin{equation}
\begin{aligned}
\label{e:domains}
&\Omega^{+}=\{r_{0}\in [0,R_{0}),\theta_{0} \in [0,\pi] ,\varphi_{0} \in(0,2\pi]\} \setminus \cup_{\a}\overline{\Omega}_{\a}\\
&\Omega^{-}=\{r_{0}\in (R_{0},\infty),\theta_{0} \in [0,\pi] ,\varphi_{0} \in(0,2\pi]\}
\end{aligned}
\end{equation}
where $\Omega_{1}= \{|\bmr_{0} |< R_{S}\}$ denotes the interior of the PS core 
and $\Omega_{\a} = \{|\bmr_{\a}|=|\bmr_{0} - \LB_{\a}|< a\}$, 
$\a=2,3,\dots,N+1$, denote the interior of the $\a$-th nanosphere.
The reactant diffuses with diffusion coefficients $D_{i}$ and $D_{o}$ inside the microgel 
shell and in the bulk, respectively.  In general one can assume $D_{i} < D_{o}$ due to obstructed 
or hindered diffusion in the microgel~\cite{Cukier:1984ve}.\\
%
%
\subsection{Steady-state boundary-value problem}
%
Let  $\rho_{B}$ denote the bulk density of reactants and let us introduce the time-dependent 
normalized density $u(\bmr,t) = \rho(\bmr,t)/\rho_{B}$. 
We assume  that the system relaxation time for the diffusive flux of 
$B$ particles (the reactants), $t_D\simeq \left( R_{0}-R_{S}\right) ^2/D_{i}$, is small enough to
neglect time-dependent effects. Hence, in the absence of external forces,
the diffusion of reactants with normalized number density  
$u(\bmr)$  is described by the steady-state diffusion equation
\begin{equation}
\nabla \cdot \left[ D(\bmr)\nabla u(\bmr)\right] =0 \qquad 
\mbox{in}\text{ }\Omega =\Omega ^{+}\cup \Omega ^{-}
\label{sp1}
\end{equation}
with 
\begin{equation}
D(\mathbf{r})=\left\{ 
\begin{array}{ll}
D_i & \mbox{in}\quad \Omega ^{+} \quad (\mbox{microgel})\\
D_o & \mbox{in}\quad \Omega ^{-} \quad (\mbox{bulk})
\end{array}
\right. 
\end{equation}
and which should be solved with the customary bulk boundary condition
\begin{equation}
\label{e:contbulk}
\lim_{|\bmr|\to\infty} u(\bmr) = 1
\end{equation}
It is well known from the general theory of partial differential equations
that the classical solution (twice continuously differentiable in $\Omega$
and continuous on $\overline{\Omega} \equiv \Omega \cup \partial \Omega_0$) 
of the stationary diffusion equation~\eqref{sp1}
does not exist in the whole domain $\Omega $~\cite{Ladyzhenskaya:1968dz}. 
Therefore one should consider the function 
\begin{equation}
u(\bmr)=\left\{ 
\begin{array}{ll}
u^{+}(\bmr) & \mbox{in}\quad \Omega^{+} \quad (\mbox{microgel}) \\
u^{-}(\bmr) & \mbox{in}\quad \Omega^{-} \quad (\mbox{bulk})
\end{array}
\right.\label{sp3}
\end{equation}
Accordingly, we should impose a condition for the 
substrate concentration field at the bulk/microgel interface,
$\partial \Omega_{0} \equiv \{r_{0}=R_{0}\}$.
It has been demonstrated recently that a key factor controlling 
the overall reaction rate  is the transfer free-energy jump $\Delta G$, 
a quantity that describes the partitioning of the reactant in the microgel 
versus bulk~\cite{Angioletti-Uberti2015}. 
For a single nanocatalyst at the nanoreactor center, a free-energy jump at the solvent-microgel 
interface can be accounted for through a modified reactant density in the microgel, namely 
$\rho \to \rho \exp(-\beta \Delta G)$ when crossing 
the bulk/microgel interface. This is also the case for many catalysts
in the infinite dilution limit. Here we assume that such description is a valid approximation for 
realistic nanoreactors, where the nanocatalyst packing fraction is
indeed very small, as discussed in depth later.  Accordingly, we require 
\begin{equation}
       \left.
           \left(
              u^{+} - 
              \lambda u^{-} 
           \right)
       \right|_{\partial \Omega_{0}} = 0 \label{e:contu}
\end{equation}
where $\lambda = \exp(-\beta \Delta G)$, $\beta = 1/k_BT$ being the inverse temperature. 
Furthermore, the following continuity condition
for the local diffusion fluxes should also hold at the bulk/microgel interface
\begin{equation}
\left.
  \left(
     \frac{\partial u^-}{\partial r_{0}} - 
     \zeta \, \frac{\partial u^+}{\partial r_{0}}
  \right)
\right|_{\partial \Omega_{0}} = 0
\label{e:contf}
\end{equation}
where we have introduced the diffusion anisotropy parameter
\begin{equation}
\label{e:defzeta}
\zeta = \frac{D_{i}}{D_{o}}
\end{equation}
Finally, reflecting boundary conditions should hold at the surface of the inert 
PS core, {\em i.e.} 
\begin{equation}
\left.\frac{\partial u^+}{\partial r_{0}}\right|_{r_{0}=R_{S}} = 0 
\label{e:reflPS}
\end{equation}
%
%
%
\subsection{The reaction rate constant}
%

We are interested in the pseudo-first-order irreversible diffusion-influenced 
reaction between the $N$ nano-catalysts $C$ encapsulated in the microgel and reactants $B$ 
freely diffusing in the bulk and in the microgel
\begin{equation}
\label{e:reaction}
C+B 
\xrightleftharpoons[k_{-D}]{k_{D}} 
C\cdot B
\xrightarrow[]{k^\ast}
C+P 
\end{equation}
where $C\cdot B$ denotes the so-called {\em encounter complex}, 
$k_D$ and $k_{-D}$ are the association and dissociation 
diffusive rate constants, respectively, and  $k^\ast$ is the surface
rate constant of the chemical reaction occurring at the reactive catalysts' boundaries. 
Reactions of the kind~\eqref{e:reaction} are customary dealt with 
by enforcing radiation boundary conditions (also known as Robin boundary conditions) at the reaction surfaces
$\de \Omega_\a$, $\a=2,3,\dots,N+1$~\cite{Collins:1949aa}, {\em i.e.} 
\begin{equation}
\left[ 4\pi a^2 D_{i} \frac{\de u^{+}}{\de r_{\a}} - k^\ast u^{+}\right]_{\partial \Omega_\a}=0 
\qquad \a=2,3,\dots,N+1
\label{e:contrad}
\end{equation}
Thus, we can consider that the nanoreactors effectively act as sinks of
infinite capacity according to the pseudo-first-order reaction scheme 
\begin{equation}
\label{e:reactionred}
C+B
\xrightarrow[]{k}
C+P 
\end{equation}
where the forward diffusion-influenced rate constant $k$ ({\em i.e.} the 
equivalent of the measured rate constant $k_{\rm obs}$~\cite{Angioletti-Uberti2015}) 
is defined by the formula 
\begin{equation}
\label{e:rate}
k=\sum_{\a=2}^{N+1} \, \int\limits_{\partial \Omega _\a}
                  \left. D_{i} \frac{\de u^{+}}{\de r_{\a}} \right|_{\partial \Omega _\a} dS 
\end{equation}
Using this rate constant one can approximately 
describe the kinetics of the effective reaction~\eqref{e:reactionred} as
\begin{equation}
c_B\left( t\right) = c_{B}(0) \exp \left( -k \, c \, t\right)
\label{sp9a}
\end{equation}
where $c = const$ is the  volume concentration of nanocatalysts within the microgel and
$c_{B}(t)$ is the time-dependent effective bulk concentration of ligands. 
We stress that our schematization of the problem holds under the {\em excess reactant} 
condition $c\ll \rho_{B}$.  Our goal is to compute the rate constant $k$ defined in 
Eq.~\eqref{e:rate}.\\
\indent Equation~\eqref{sp1} with the boundary 
conditions~\eqref{e:contbulk},~\eqref{e:contu},~\eqref{e:contf},~\eqref{e:reflPS} and~\eqref{e:contrad}  
completely specify our mathematical problem. It is expedient in the following to
use the dimensionless spatial variables $\xi_{0} =r_{0}/R_{0}$, $\xi_{1} =r_{0}/R_{S}$ and $\xi_{\a} =r_{\a}/a$ for 
$\a=2,3,\dots,N+1$. Hence, our problem can be cast in the following form
\begin{subequations}
\label{e:BVP}
\begin{align}
&\nabla^2 u^{\pm} = 0 \quad \mbox{in} \ \Omega^{\pm} \label{e:BVPa}\\
&\left.\left(
  \frac{\partial u^+}{\partial \xi_{\a}} - h u^+ 
  \right)\right|_{\partial \Omega_{\a}}= 0 \quad \a=2,3,\dots,N+1
\label{e:BVPb}\\
&\lim_{\xi_{0}\to\infty} u^-(\xi_{0}) = 1\label{e:BVPc}\\
&\left. \frac{\partial u^{+}}{\partial \xi_{1}} \right|_{\xi_{1}=1}=0\label{e:BVPd}\\
& \left. \left(
     u^{+} - \lambda u^{-}
  \right)\right|_{\partial \Omega_{0}} = 0 
\label{e:BVPe}\\
& \left. \left(
     \zeta \, \frac{\partial u^+}{\partial \xi_{0}} - 
              \frac{\partial u^-}{\partial \xi_{0}}
  \right)\right|_{\partial \Omega_{0}} = 0
\label{e:BVPf}
\end{align}
\end{subequations}
The parameter 
\begin{equation}
\label{e:acca}
h = \frac{k^\ast}{4\pi D_{i} a} \equiv \frac{k^\ast}{k_{S}^+} 
\end{equation}
gauges the {\em character} of the reaction. Here we have introduced the Smoluchowski 
rate constant for a nanocatalyst embedded in the microgel, $k_{S}^+ = 4\pi D_{i}a$.
The limit~$h\to\infty$ corresponds to 
considering the boundaries $\de\Omega_{\a}$ as perfectly absorbing sinks. 
In this case the reaction~\eqref{e:reaction} 
becomes {\em diffusion-limited}, as the chemical conversion from the encounter complex $C\cdot B$ to 
the product $P$ becomes infinitely fast with respect to the diffusive step leading to the 
formation of $C\cdot B$. Otherwise, for $h\ll 1$, the chemical conversion step 
is slow enough compared to diffusion, which  makes the reaction overall reaction-limited. 
%

%
\section{Exact solution of the problem and approximate analytical treatment\label{sec:sol}}
%

We look for solutions for the stationary density of reactants
in the bulk and in the microgel as linear combinations of regular and irregular harmonics. 
Given the multi-connected structure of the boundary manifold $\cup_{\a}\Omega_{\a}$, 
we must consider as many {\em local} Cartesian reference frames as there are non-concentric 
boundaries.   Thus, we can look for solutions in the form
\begin{subequations}
\label{e:upm}
\begin{align}
&u^{+}(\bmr) = \sum_{\ell=0}^{\infty}\sum_{m=-\ell}^{\ell}A_{m\ell}\,\xi_0^{\ell}\,Y_{m\ell}(\bmr_{0}) + 
               \sum_{\alpha=1}^{N+1}
               \sum_{\ell=0}^{\infty}\sum_{m=-\ell}^{\ell}
                    \frac{B^{\a}_{m\ell}}{\xi_\alpha^{\ell+1}} \,Y_{m\ell}(\bmr_{\a}) \\
&u^{-}(\bmr) = 1 + \sum_{\ell=0}^{\infty}\sum_{m=-\ell}^{\ell}E_{m\ell}\,\xi_0^{-\ell-1}\,Y_{m\ell}(\bmr_{0})
\end{align}
\end{subequations}
where $Y_{mn}(\bmr)$ are spherical harmonics, $\xi_{0} = r_{0}/R_{0}$, $\xi_{1}=r_{0}/R_{S}$, 
$\xi_{\a} = r_{\a}/a$ for $\a=2,3,\dots,N$
and $A_{mn},B^{\a}_{mn}$ and $E_{mn}$ are $N+3$ infinite-dimensional sets of unknown coefficients
that can be determined by imposing the boundary conditions~\eqref{e:BVPb} and~\eqref{e:BVPd} 
and the pseudo-continuity conditions at the microgel-solvent interface, eqs~\eqref{e:BVPe}
and~\eqref{e:BVPf}. This can be done straightforwardly using 
known addition theorems for spherical harmonics, which results in an infinite-dimensional 
linear system of equations for the unknown coefficients (see appendix~\ref{app:A} for the details).
Furthermore, making use of known properties of solid spherical harmonics,
it is easy to see that the rate constant defined by Eq.~\eqref{e:rate} is simply given  by
\begin{equation}
\label{e:rateB0}
k = - k^+_{S} \sum_{\a=2}^{N+1} B_{00}^{\alpha}
\end{equation}
As shown in the appendix~\ref{app:A}, the exact solution to the steady-state 
problem~\eqref{e:BVP} can be worked out in principle to any desired precision by keeping an appropriate 
number of multipoles. Remarkably, a simple yet accurate {\em analytical} expression can be easily obtained in the 
monopole approximation (MOA), which corresponds to keeping only the 
$\ell=0$ term in the multipole expansions~\eqref{e:upm}~\cite{Deutch:1976aa,Felderhof:1976aa}. 
In particular, it is interesting  to compute the rate 
normalized to the Smoluchowski rate of an isolated sink of the same size as the whole 
nanoreactor in the bulk, {\em i.e.} $k_S^- = 4\pi D_0 R_0$. We obtain 
(see appendix~\ref{app:B} for the details)
\begin{equation}
\label{e:ratef}
\frac{k}{k_S^-} = Nk^\ast 
                  \left( 
                          \frac{a}{R_0}
                  \right) 
                  \frac{\ds\zeta e^{-\beta \Delta G}}
                       {\ds k_S^+ + k^\ast\bigg[
                                            1 + (N-1)\left\langle 
                                                       \frac{a}{L_{\alpha\beta}}
                                                     \right\rangle - 
                                                 \frac{\ds Na}{\ds R_0}
                                                 \left(
                                                    1 - \zeta e^{-\beta \Delta G} 
                                                 \right)
                                      \bigg]}
\end{equation} 
where we recall that $\zeta = D_i/D_0$ and $k_S^+ = 4\pi D_i a$. 
This is the key analytical  result derived in this work, that can be readily employed to predict and optimize
the geometry and activity of typical core-shell nanoreactors.
The quantity $\langle a/L_{\alpha\beta} \rangle$ stands for the average
inverse inter-catalyst separation. This can be computed analytically under the 
reasonable assumption that spatial correlations in the catalysts configurations
are negligible (see appendix~\ref{app:B}), 
\begin{equation}
\label{e:averinvL}
\begin{aligned}
\left\langle
   \frac{a}{L_{\alpha\beta}}
\right\rangle &= \frac{2(1-\ve)^5 - 5 (1-\ve)^2(\gamma+\ve)^3 + 3(\gamma+\ve)^5}
                       {(1-\ve)^6  - 2 (1-\ve)^3(\gamma+\ve)^3 +  (\gamma+\ve)^6}  
                  \left( 
                      \frac{3a}{5R_0} 
                  \right)      \\ 
                  &:= \ve \, C (\ve,\gamma) 
\end{aligned}                  
\end{equation}  
where $\gamma=R_S/R_0$ denotes the fraction of the nanoreactor size occupied by the PS core 
and $\ve=a/R_0$ is the non-dimensional size of each catalyst. 
We see that, since $\ve \ll 1$, one has $ 1 + \ve/3 \lesssim C \lesssim 6(1+\ve)/5 $,
{\em i.e.},  $C$ is of the order of unity, $ 1.005 \lesssim C \lesssim 1.217$
(taking $\ve \approx 0.0146$ from experiments~\cite{Mei:2007vn}). \\
\indent In the limit of vanishing surface reactivity of the embedded nano-catalysts
it is immediate to show from eq.~\eqref{e:ratef}  that 
\begin{equation}
\label{e:ratesmallks}
k \simeq Nk^\ast e^{-\beta \Delta G} + \mathcal{O}[(k^\ast/k_S^+)^2]
\end{equation}
We see that, if the surface substrate conversion rate constant is weak, this becomes the rate-limiting 
step for the overall rate of the nanoreactor, irrespective of the geometrical features of 
the assembly and of the mobility properties of the hydrogel shell. 
In this case, it becomes crucial to control the transfer free energy jump 
to tune the rate of the composite nanoreactor.
\indent Conversely, if the catalytic action exerted by the metal nanoparticles encapsulated 
in the microgel is fast with respect to diffusion, {\em i.e.}  $k^\ast \gg k_S^+$, 
expression~\eqref{e:ratef} can be simplified by taking the limit $k^\ast \to \infty$.
This yields the expression for the fully diffusion-controlled rate 
\begin{equation}
\label{e:ratefDL}
\frac{k}{k_S^-} = \frac{\ds N\ve\,\zeta e^{-\beta \Delta G}}
                       {\ds  1 + (N-1)\ve\,C(\ve,\gamma) - 
                                            N\ve
                                            \left(
                                               1 - \zeta e^{-\beta \Delta G} 
                                            \right)}
\end{equation} 
which we will discuss in depth in the following section. Note that for $N=1$ Eq.~\eqref{e:ratefDL} 
coincides with the solution of the Debye-Smoluchowski problem~\cite{debye42} for a single 
perfectly absorbing sink located at the center of the shell, with 
$G(r)=\{\Delta G \ \mbox{for} \ a<r\leq R_0 \,| \,0 \ \mbox{for} \ r>R_0\}$.\\
\indent Formulas~\eqref{e:ratef} and~\eqref{e:ratefDL} have been derived in the monopole approximation,
which means that any reflecting boundaries in the problem are not taken into account. 
Therefore, these should be used to approximate the rate constant of a composite nanoreactor for small 
to moderate sizes of the central  PS core. 
In the following sections, we provide a thorough characterization of the rate constant of 
a composite core-shell nanoreactor,  computed exactly  by solving Eqs.~\eqref{sistemananoridotto},
and we compare it to the approximate MOA analytical expression~\eqref{e:ratefDL} in the 
physically relevant diffusion-limited regime ($k^\ast \to \infty$).

%
\section{The diffusion-controlled regime\label{sec:analytic}}
%
%
We now discuss in more detail the essential features of the diffusion-controlled rate in Eq.~\eqref{e:ratefDL}. 
In the monopole approximation, valid for small to intermediate sizes of the central reflecting (inert) core, 
the role of the latter only enters 
indirectly through the spatial average $\left\langle {a}/{L_{\alpha\beta}} \right\rangle = Ca/R_0$, 
with $C\simeq 1$. 
In the swollen configuration, the central core does not occupy a large fraction of the 
overall nanoreactor volume, with  $\gamma \approx 0.3$ (as taken from the experiments reported in 
Ref.~\cite{Mei:2007vn}). Hence, in this regime
we expect that the exact size of the core should not play a significant role for the 
diffusion-controlled rate for relevant values of the physical parameters,  {\em i.e.}, 
(weak) attraction to the hydrogel $\Delta G < 0$ and decreased internal diffusion $\zeta <1$. \\
%
%
\begin{figure*}[t!]
\centering
\begin{tabular}{cc}
\resizebox{8.5 truecm}{!}{\includegraphics[clip]{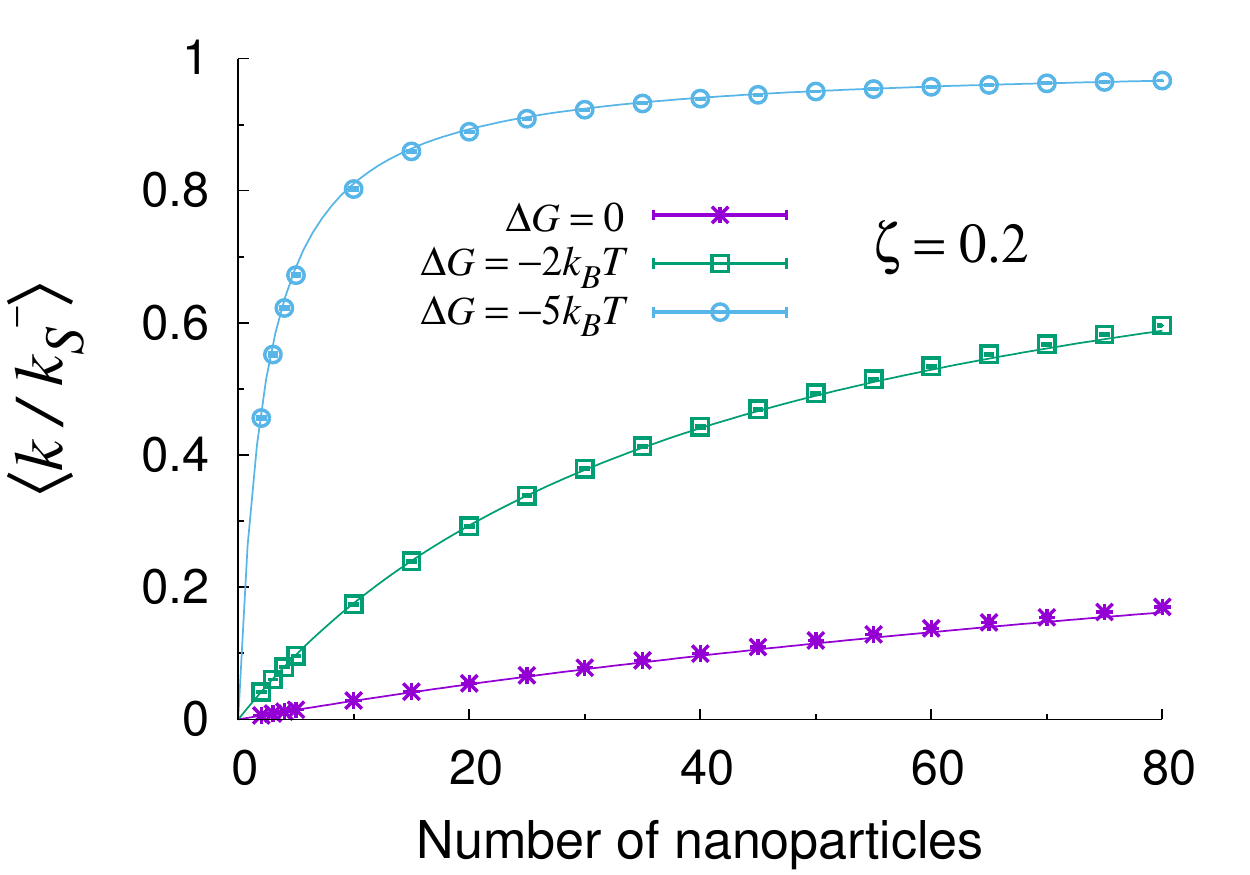}} &
\resizebox{8.5 truecm}{!}{\includegraphics[clip]{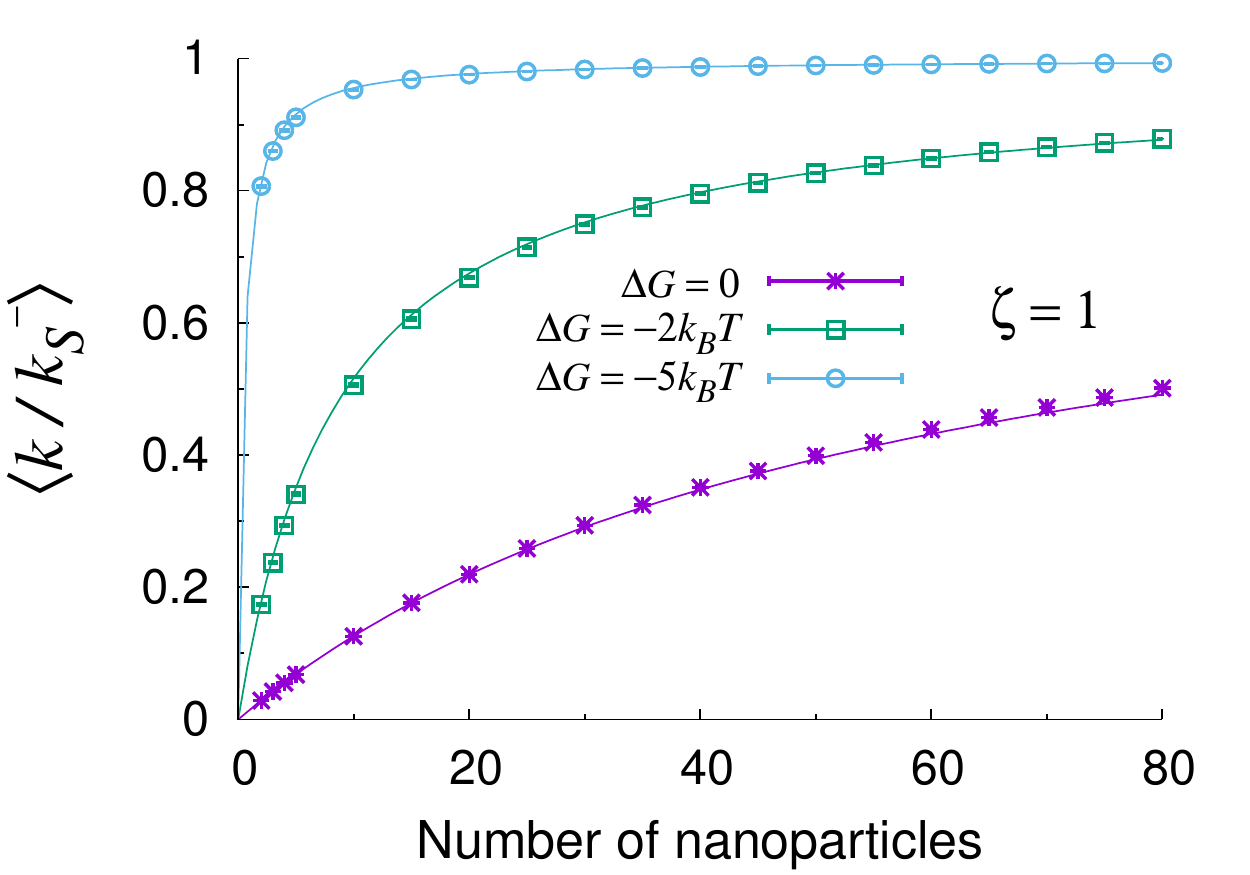}} 
\end{tabular}
\caption{Diffusion-controlled rate constant of a core-shell nanoreactor normalized to the Smoluchowski 
rate constant of a perfect sink of the same size, $k_S^- = 4\pi D_o R_0$ versus
the number $N$ of encapsulated nanoparticles (NP). Symbols are the exact solution of Eqs.~\eqref{sistemananoridotto}
(relative accuracy of $TOL=5\times 10^{-4}$). Each point is an average over 10 independent 
configurations of the NPs (error bars smaller than the symbol size), while the solid lines 
are plots of the monopole approximation, Eq.~\eqref{e:ratefDL}, for the corresponding choice of parameters. 
The plots refer to $R_S/R_0=0.353$ and $a/R_0 = 0.0146$. 
\label{f:rates}}
\end{figure*}
%
\indent In Fig.~\ref{f:rates} we compare the approximate expression~\eqref{e:ratefDL} 
to the exact solution of Eqs.~\eqref{sistemananoridotto} 
for two different mobility ratios $\zeta = 0.2$ and 1.0 and 
core size  $\gamma=0.353$ (as in previous experiments~\cite{Mei:2007vn}).  
The NP size is held fixed as $\ve = 0.0146$ as also provided from experiments.  
It is apparent  that the analytical treatment is  remarkably 
accurate in these conditions and deviates from the exact solution by less than one percent in the worst cases. 
This proves that our analytical treatment provides a reliable tool for realistic values
of the physico-chemical and geometrical parameters.
A comparison of different values of the substrate mobility within the gel ($\zeta$), 
clearly highlights that all rates are higher when the ligand 
is more mobile within the gel shell. Concerning the overall 
form of the curves, one can see that the initial linear rise of the rates is followed by a saturation
at large values of $N$. The approach to saturation is slow for small values of 
$\Delta G$, but begins markedly earlier ({\em i.e.} for smaller $N$) if the sorption free 
energy reaches values as small as a few $k_BT$, the thermal energy. 
Hence, as discussed already previously, a decisive factor in the 
design of optimized nanoreactors must be clearly the tuning of the reactant-hydrogel 
interaction towards attraction.
We note that free energy gains $\Delta G$ of the order of a few $k_BT$ seem utterly 
realistic for small hydrophilic substrates, such as nitrobenzol and nitrophenyl. 
For comparison, a reasonable upper bound could be estimated as the 
free energy jump of about 7 kT reported for the sorption of a protein 
into a hydrophilic network~\cite{Yigit:2012aa}. These estimates are also consistent with partitioning 
data of small molecules, such as acetaminophen, into PNIPAM~\cite{Palasis:1992aa}, where 
the transfer free energy can be estimated as $\Delta G \simeq -k_BT \ln K$, where $K$ 
is the partitioning coefficient.

%
%
\subsection{Optimizing the number of nanocatalysts}
As we see from Eq.~\eqref{e:ratefDL2} the maximum achievable rate is $k = k_S^- = 4\pi D_0 R_0$, 
that is, the Smoluchowski rate of a sink of size equal to that of the total nanoreactor, {\em i.e.} 
the nanoreactor should be big for high activity. In the limit of small NP to nanoreactor size 
ratio, $\ve\ll 1$, Eq.~\eqref{e:ratefDL} can be simplified to the following form
\begin{equation}
\label{e:ratefDL2}
\frac{k}{k_S^-} = 
                  \frac{\ds N\ve\,\zeta e^{-\beta \Delta G}}
                       {\ds  1 + N\ve\,\zeta e^{-\beta \Delta G}                                              }.
\end{equation} 
Let us recall the important parameters, that is, the NP to nanoreactor size ratio $0 < \ve=a/R_0 \ll 1 $, 
the number of NPs $N$, the scaled reactant mobility inside the shell $0 < \zeta = D_i/D_0 \lesssim 1 $,  
and finally  the transfer free energy  change $\Delta G$ for the reactants upon entering the hydrogel.  
Clearly, if the mobility vanishes, $\zeta \ll 1$ or the   free energy jump $\Delta G \gg k_BT$ 
is substantially repulsive, the reaction is significantly slowed down. However, in realistic systems 
the mobility will be certainly slowed down to some extent but not vanish. $\Delta G$ may be even negative 
(attractive) if  the reactant interacts favorably with the polymer as found for rather hydrophobic reactants and 
collapsed PNIPAM-based hydrogels~\cite{Wu2012,Angioletti-Uberti2015}. 
Since $\Delta G$ enters Eq.~\eqref{e:ratefDL2} exponentially,  substantial effects are expected 
following small changes in the interaction.  Together with $\Delta G$, clearly the number of NPs and their 
size ratio with respect to the total nanoreactor size are the key quantities
to tune. To save resources $N$ should be small but large enough to warrant a high catalytic activity.\\
\indent The behavior of Eq.~\eqref{e:ratefDL2} resembles a Langmuir-binding isotherm form. 
The rate as a function of $N$ initially rises linearly with a slope $\ve\,\zeta\exp(-\beta \Delta G)$ 
and finally saturates to the maximum rate $k = k_S^-$ \, for large 
values of $N$.  For a single NP, $N=1$ and not too attractive transfer free energy, 
we recover essentially the result for a yolk-shell 
nanoreactor $k \simeq 4\pi D_i a\exp(-\beta \Delta G)$, where a single NP is embedded in the center of a 
spherical hydrogel,  apart from a slight modification of the target size, which is not $a$ for the 
yolk-shell but $R_i$, the radius of the interior hollow confinement~\cite{Wu2012,Angioletti-Uberti2015}. \\
\indent It is instructive to define an efficieny factor $f = k/k_S^-$ between 0 and 100 $\%$, 
that quantifies the desired target  efficiency of the nanoreactor. 
Solving Eq.~\eqref{e:ratefDL2} for $N$,  we find 
\begin{equation}
\label{e:thumb}
N_f =  \left( 
           \frac{e^{\beta\Delta G}}{\ve\,\zeta}
       \right)\frac{f}{1-f}
\end{equation}
that is, for a fixed efficiency, the NP number needed to maintain it changes exponentially 
with the transfer free energy change. 
As a numerical example, let us assume reasonable values of $\zeta = 0.2$, $\ve=0.01$, and $\beta\Delta G = -1$. 
To obtain  an efficiency of 50 $\%$, $N_f=184$ nanoparticle catalysts would be needed. 
If $\beta\Delta G = -2$, the number wold drop of a factor $1/e$ to $N_f=68$.  
For such values of $N$ the MOA is an excellent approximation (see Fig.~\ref{f:rates}), 
which makes our treatment self-consistent and sound.
Note that $N_f$ does not scale with the catalyst {\em surface}, as one might naively expect, 
rather it decreases {\em linearly} with the catalyst size. \\
\indent Formula~\eqref{e:thumb} provides a simple rule of thumb for optimizing the design and
synthesis of core-shell nanoreactors for small to intermediate values of the core size.  
As an example,  if one aims at 50 $\%$ 
efficiency for a relatively neutral hydrogel chemical environment ($\Delta G = 0$),
where the mobility of the substrate is not significantly reduced ($\zeta=1$), 
one needs to employ $N_f = 1/\ve = R_0/a$ nanoparticles. For  $\ve=0.01$ that would be $N_f=100$.
In the case of a polymer matrix in physical-chemical conditions leading to a reduced  
mobility ({\em e.g.} $\zeta=0.2$), one would need five times more NPs for $\Delta G=0$,
but about the same number for $\beta \Delta G \simeq -1.6$. This clearly illustrates 
how the {\em performance} of a composite core-shell nanoreactor is non-trivially shaped by the combined 
action of the physical chemical properties of the hydrogel shell matrix, such as the bulk solvent-microgel
transfer free energy jump and changes in translational mobility of the substrate molecules.

%
\begin{figure*}[t!]
\centering
\begin{tabular}{cc}
\resizebox{8.5 truecm}{!}{\includegraphics[clip]{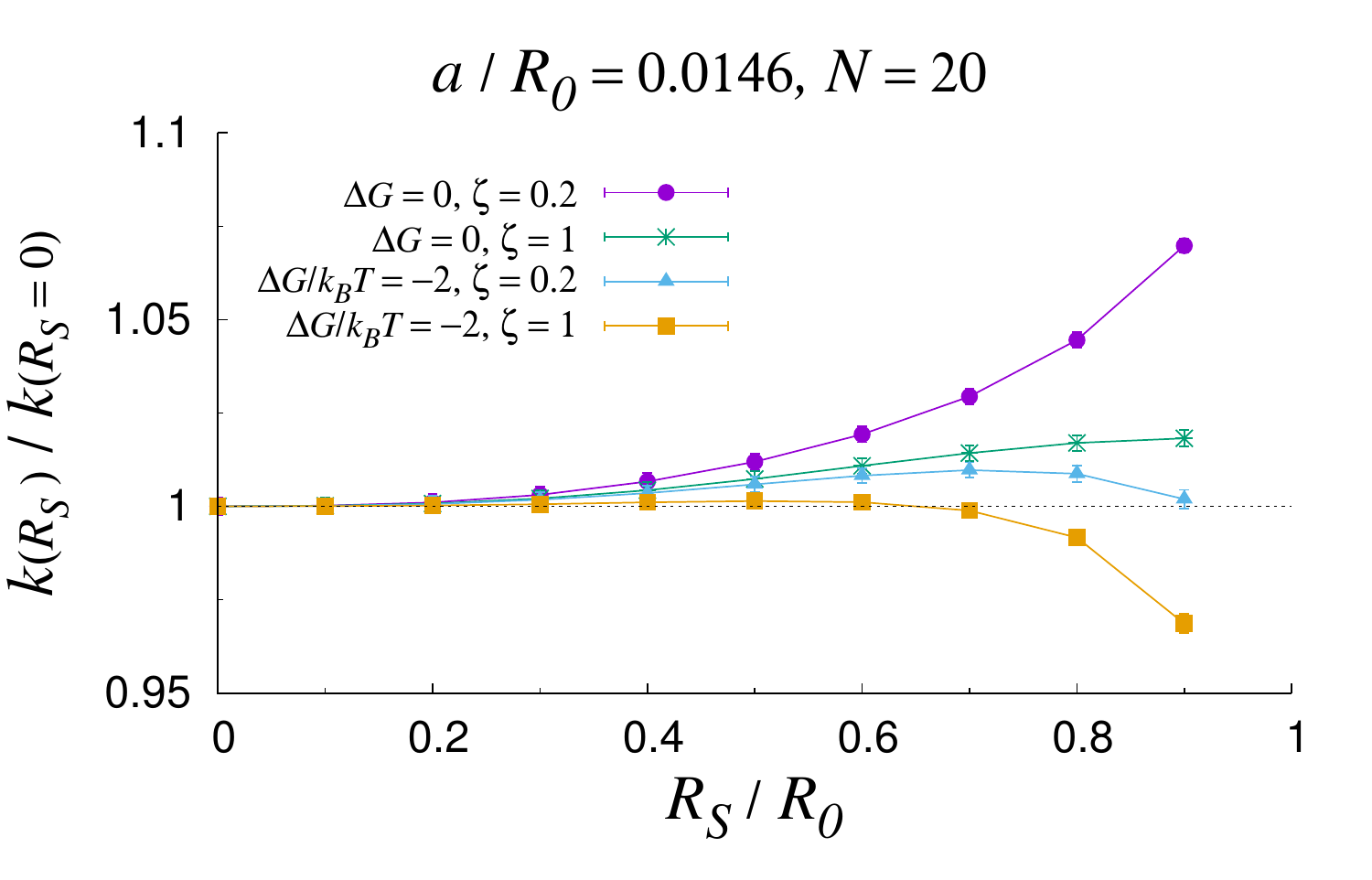}}  &
\resizebox{8.5 truecm}{!}{\includegraphics[clip]{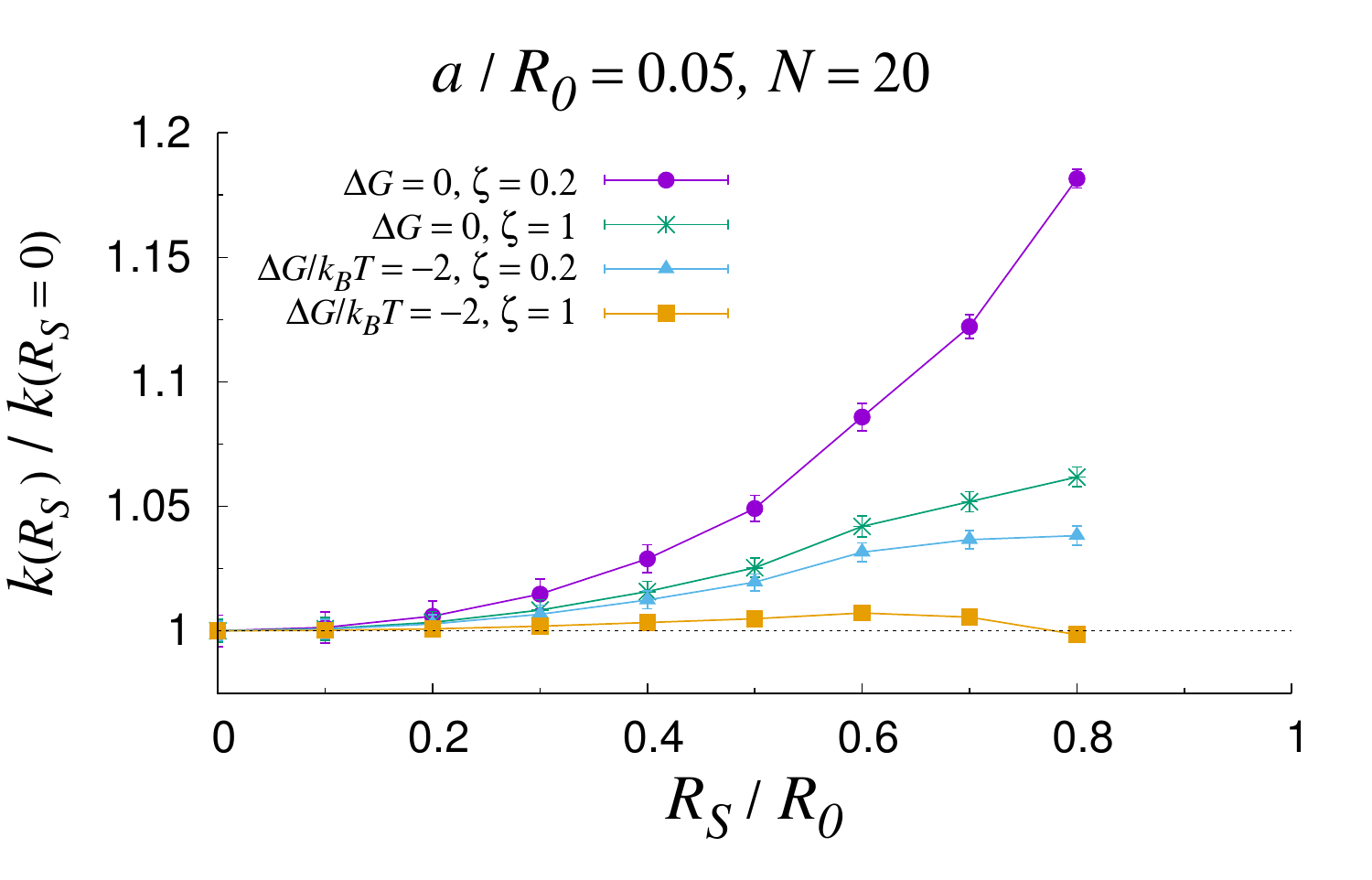}}    \\
\resizebox{8.5 truecm}{!}{\includegraphics[clip]{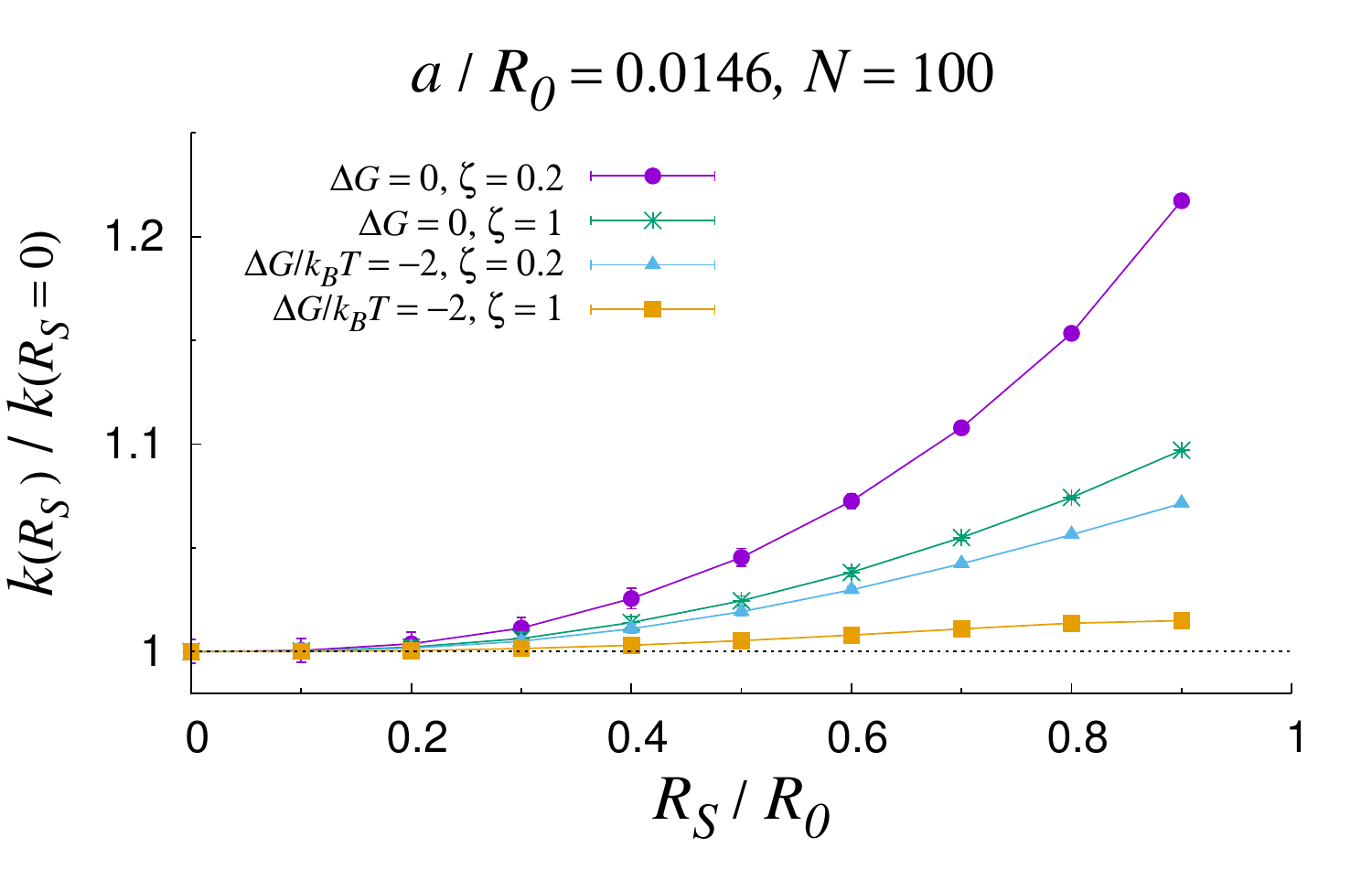}} &
\resizebox{8.5 truecm}{!}{\includegraphics[clip]{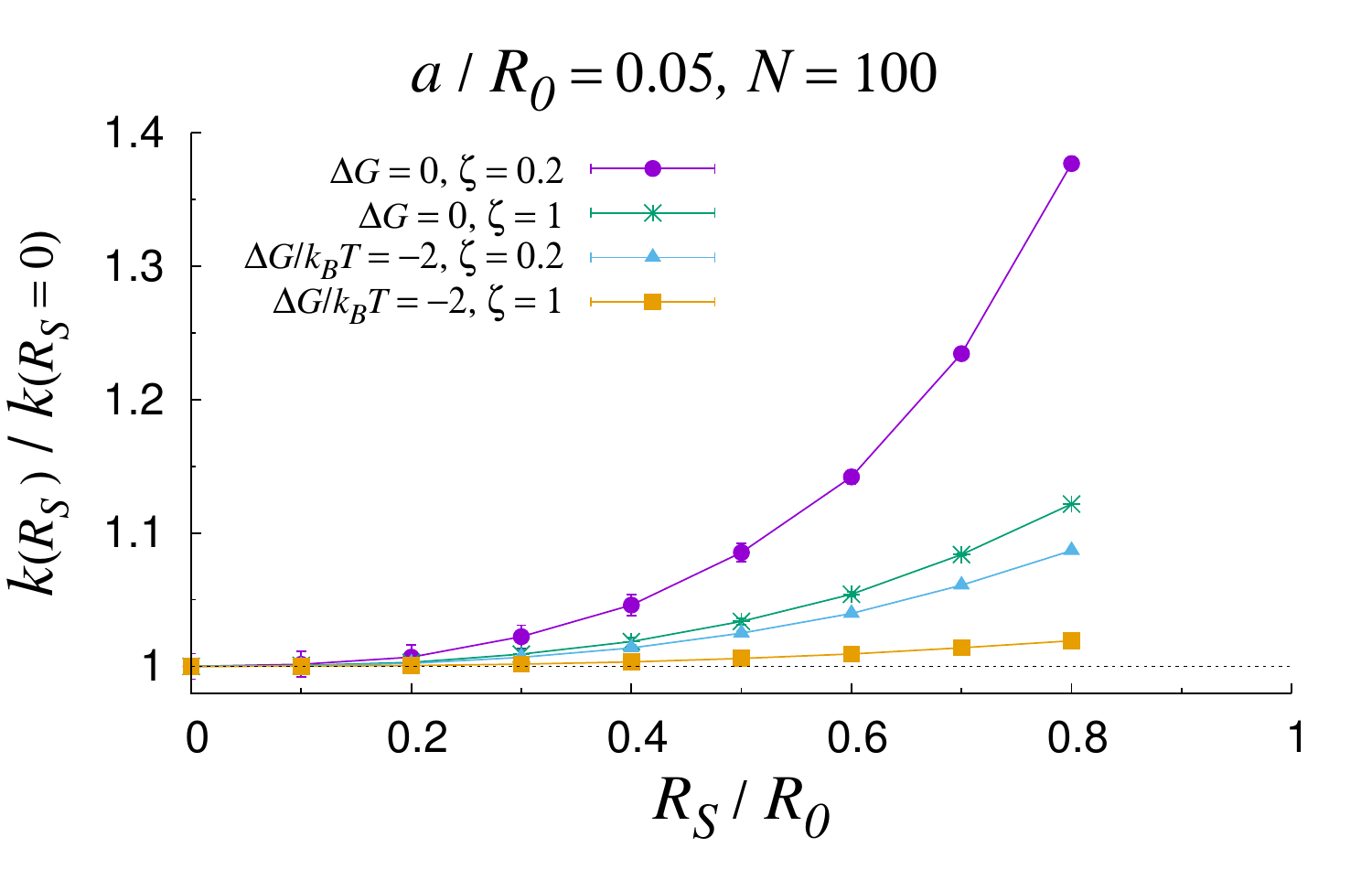}} 
\end{tabular}
\caption{Diffusion-controlled rate constant of a core-shell nanoreactor as a function of 
the PS core size for different values of the physico-chemical parameters and nanocatalyst
loading number.  
\label{f:versusRS}}
\end{figure*}
%

\subsection{The role of the core size}

In certain configurations, such as in the shrunk phase of thermosensitive core-shell 
nanoreactors past the lower critical solution temperature (LCST)~\cite{Lu2011}, the core size can 
become comparable to the overall size of the nanoreactor. In these circumstances, the MOA breaks 
down and the full solution should be used instead. Fig.~\ref{f:versusRS} reports an analysis 
of the rate constant as the core sized is varied for different values of the geometrical and
physico-chemical parameters. As a first observation, the plots confirm and substantiate the 
discussion laid out in the previous section, as it can be appreciated that the core
size does not influence the overall rate until $\gamma=R_S/R_0\lesssim 0.4$.
More generally, one can recognize that the rate constant tends to increase as the shell shrinks
(increasing values of $R_S/R_0$). The only exception is for low $N$ and attractive 
transfer free energy, where a non-monotonic trend is observed (top left panel). This is 
a typical screening effect~\cite{Galanti:2015aa}, which originates from the subtle 
interplay between diffusive interactions among the nanocatalysts and individual 
screening due the reflecting PS core. 
It turns out that the transfer free energy is the  prime parameter that controls the increase in the rate as the 
PS core size increases. The more attractive the transfer free energy, the less marked the increase. 
Interestingly, at fixed values of $\Delta G$, the less mobile the substrate in the shell,
the more marked the rate boosting effect of the shell shrinking.
Importantly, it is apparent from the plots reported in Fig.~\ref{f:versusRS} that 
the role of the core size is reduced for loading number of the order of a few tens
and small size of the nanocatalysts.
All in all, these results confirm the complex intertwining of the structural, geometrical and physico-chemical 
features underlying the overall catalytic activity of core-shell nanoreactors.

%
%

\section{Concluding remarks\label{sec:summary}}

In this paper we have developed a detailed theory to compute the  total reaction rate of core-shell 
nanoreactors with multiple catalysts embedded in the shell. The theory is utterly general and 
allows one to compute the overall reaction rate to any desired accuracy for ($i$) given configuration, 
dimension and surface reactivity  of the encapsulated nanocatalysts, ($ii$) size of the core and the shell, 
($iii$) substrate mobility in the bulk and in the shell and ($iv$) transfer free-energy jump for substrate molecules. 
Furthermore,  we computed analytical expressions  in the monopole approximation that provide 
an excellent interpolation of the exact solution for small to intermediate sizes of the 
central core in the physically relevant range of parameters,
{\em i.e.} small size and high dilution of the nanocatalysts. Our formulas supply ready-to-use 
simple tools that can be employed to interpret and optimize the activity of experimentally realizable 
nanoreactor systems.  This shall be particularly useful to estimate the optimal number of embedded NPs, 
that should reflect a compromise between a resource-friendly design and the highest possible catalytic output. 
Our analytical treatment predicts an optimal number of NPs given by the following expression
\begin{equation}
\label{e:thumbf}
N_f = \frac{f}{1-f} \frac{R_0}{a} 
                     \left( 
                            \frac{D_o}{D_i}            
                     \right) e^{\beta\Delta G}
\end{equation}
where $f\in[0,1]$ is the desired efficiency, $a$ and $R_0$ are the NP and overall nanoreactor
sizes, respectively, and $D_i$,$D_o$ are the substrate mobility in the shell and in the 
bulk, respectively. For realistic values of these parameters, one gets $N_{1/2}$ of the 
order of hundreds, a value for which the monopole approximation is still in excellent
agreement with the exact solution for core sizes such that $R_S/R_0 \lesssim 0.4$. 
As discussed already previously, eq.~\eqref{e:thumbf} makes it clear that a decisive factor 
in the design of optimized hydrogel-based  nanoreactors must be  the tuning of the reactant-hydrogel 
interaction towards attraction ($\Delta G <0$) for a specific reaction (or mix of reactions).  
Furthermore, as hydrogel that cause strongly reduced substrate mobility also demand more
NPs to achieve high efficiency ($N_f \propto D_o/D_i$), the choice of the shell hydrogel should 
be made so as to privilege smooth longer-ranged interactions 
(like electrostatic, hydrophobic, or dispersion)  with respect to short-ranged ones (like H-bonds),
in order to avoid too sticky interactions that would slow down the reactant mobility 
substantially due to activated hopping.\\
\indent Our analytical treatment breaks down if one   
wishes to push the nanoreactor performances towards full efficiency ($f\to 1$), where 
the loading number of NP increases rapidly, or in the case of larger core
sizes (of the order of the whole spherical assembly). In such cases, the diffusive interaction 
between NPs can no longer be neglected, as well as the effect of the inert PS core, 
as diffusive and screening interactions 
among the different boundaries become important. As a consequence, the full exact solution should be employed 
to  investigate the behavior of the rate and elaborate an optimal design of the composite 
nanoreactor.  Interestingly, we have  shown that, as a general situation, increasing both the core  and the 
nanocatalyst sizes either has a rather mild effect on the overall performances, or, more generally,
causes a rate-boosting effect, with an increase of the overall rate  constant of up to 40 \%
for values of the core size $R_S/R_0 \gtrsim 0.7$.

%

\section*{Acknowledgments}

S. A-U acknowledges financial support from the Beijing Municipal Government 
Innovation Center for Soft Matter Science and Engineering.
J. D. acknowledges funding by the ERC (European Research Council) 
Consolidator Grant with project number 646659--NANOREACTOR. M. G. and F. P.
would like to thank S. D. Traytak for insightful discussions. F. P.  and D. F. 
acknowledge funding by the CNRS (Centre National de la Recherche Scientifique) 
under the PICS scheme.

\appendix
\section{\label{app:A}}

\indent In order to determine the unknown coefficients in the 
expansions~\eqref{e:upm}, we have to express the solution 
in the local coordinates on every boundary (the $N+1$ spherical surfaces $\partial\Omega_{\a}$) 
and at the microgel-bulk interface $\partial \Omega_{0}$,
where we impose the pseudo-continuity conditions for the reactant density field.
This can be accomplished by using known addition theorems for spherical 
harmonics~\cite{Arfken:2005cr,Caola:2001pd}. 
After some lengthy algebra,  we obtain the following linear equations
\begin{subequations}
\begin{align}
&\frac{1}{\lambda} A_{gq} + \frac{1}{\lambda}\sum_{\beta=1}^{N}\sum_{n=0}^{q}\sum_{m=-n}^{n}B_{mn}^{\beta}
          V_{g,q}^{\beta,m,n}\mathbb I _{\{g-(q-n)\leq m \leq g+(q-n)\}}-E_{gq}=\delta_{g0}\delta_{q0}
          \label{sistemanano1}\\
&-B_{gq}^{\alpha} + \frac{(q-h_\alpha)}{(h_\alpha+q+1)}
         \sum_{n=0}^{\infty}\sum_{m=-n}^{n}\bigg(
                    A_{mn} H_{m,n}^{(\alpha,g,q)}
                    \mathbb I _{q\leq n}+\sum_{\beta=1,\beta\neq \alpha}^{N}B_{mn}^{\beta} 
                    W_{m,n}^{(\alpha,\beta,g,q)}\bigg)=0\label{sistemanano2}\\
& \zeta\left[  \sum_{\beta=1}^{N}\sum_{n=0}^{q}\sum_{m=-n}^{n}B_{mn}^{\beta}
               V_{g,q}^{\beta,m,n}\mathbb I _{\{g-(q-n)\leq m \leq g+(q-n)\}} -\frac{q}{q+1} A_{gq}
        \right] - E_{gq}=0\label{sistemanano3}
\end{align}
\end{subequations}
where $h_{1}=0$, $h_{\a}=h$ for $\a>1$ and we have introduced characteristic functions 
$\mathbb I_{m\in \mathcal{I}}=\{1 \ \text{for} \ m \in \mathcal{I}| 0 \ \text{otherwise}\}$.
Eqs.~\eqref{sistemanano1},~\eqref{sistemanano2},~\eqref{sistemanano3} hold $\forall \ q\in[0,\infty)$ with 
$\alpha=1,2,\dots,N+1$ and $g=-q,-q+1,\dots,q-1,q$.  The matrices $V,H,W$ read
\begin{subequations}
\begin{align}
& V_{g,q}^{\alpha,m,n}=
\frac{(-1)^{q-n+m-g}(q-g)!}{(n-m)!(q-n+m-g)!}\eta_{0\alpha}^{q-n}\chi_{\alpha}^{n+1} Y_{m-g,q-n}(-\LB_{\alpha}) 
\label{e:Vdef}\\
& H_{m,n}^{(\alpha,g,q)}= 
\binom{n+m}{q+g}  \chi_{\alpha}^q \eta_{0\alpha}^{n-q}Y_{m-g,n-q}(\mathbf{L}_{\alpha}) 
\label{e:Hdef}\\
& W_{m,n}^{(\alpha,\beta,g,q)} =(-1)^{q+g}\frac{(n-m+q+g)!}{(n-m)!(q+g)!}\eta_{\beta\alpha}^{-(n+q)-1} \chi_{\alpha}^{q}
\chi_{\beta}^{n+1}Y_{m-g,n+q}(\LB_{\beta\alpha}) \label{e:Wdef}
\end{align}
\end{subequations}
where $\LB_{\a\b} = \LB_{\b} - \LB_{\a}$ (according to this notation $\LB_{0\a}=\LB_{\a}$) and 
\begin{equation}
\label{e:etachi}
\eta_{\alpha\beta}=\eta_{\beta\alpha}=\frac{L_{\alpha\beta}}{R_0} \qquad \qquad 
\chi_\alpha=\frac{R_\alpha}{R_0}.
\end{equation}
Here, for the sake of coherence, we pose $R_{1} = R_{S}$ (radius of the PS core) and $R_{\a} = a$,
for $\a>1$ (radius of the nanocatalysts). The system~\eqref{sistemanano1},~\eqref{sistemanano2},~\eqref{sistemanano3}
can be expressed more conveniently by subtracting eq.~\eqref{sistemanano3}  
from eq.~\eqref{sistemanano1}, which leads to
\begin{widetext}
\begin{subequations}
\label{sistemananoridotto}
\begin{align}
&\bigg(\frac{1}{\lambda}+\frac{q}{q+1}\zeta\bigg) A_{gq} + \bigg(\frac{1}{\lambda}-\zeta\bigg)
          \sum_{\beta=1}^{N}\sum_{n=0}^{q}\sum_{m=-n}^{n}B_{mn}^{\beta}
          V_{g,q}^{\beta,m,n}\mathbb I _{\{g-(q-n)\leq m \leq g+(q-n)\}} = \delta_{g0}\delta_{q0}
          \label{sistemananoridottoa}\\
&-B_{gq}^{\alpha} + \frac{(q-h_\alpha)}{(h_\alpha+q+1)}
         \sum_{n=0}^{\infty}\sum_{m=-n}^{n}\bigg(
                    A_{mn} H_{m,n}^{(\alpha,g,q)}
                    \mathbb I _{q\leq n}+\sum_{\beta=1,\beta\neq \alpha}^{N}B_{mn}^{\beta} 
                    W_{m,n}^{(\alpha,\beta,g,q)}\bigg)=0 \label{sistemananoridottob}
\end{align}
\end{subequations}
\end{widetext}
If the multipole expansions are truncated at $N_M$ multipoles, 
the system~\eqref{sistemananoridotto}) comprises $(N+2)(N_M+1)^2$ equations, 
which can be easily solved numerically. Once the the coefficients have been determined, 
the rate constant can be  obtained from eq.~\eqref{e:rate}.
Recalling the definitions~\eqref{e:upm} and making use of known properties of solid spherical harmonics,
it is easy to see that
\begin{equation}
\label{e:rateBapp0}
k = - k^+_{S} \sum_{\a=2}^{N+1} B_{00}^{\alpha}
\end{equation}
The system to be solved has the following structure
\begin{widetext}
\[ 
\small
\left[
\begin{BMAT}(@)[4pt]{c.cccc}{c.cccc}
\bigg(\frac{1}{\lambda}+\frac{\zeta q}{q+1}\bigg) \1 & \bigg(\frac{1}{\lambda}-\zeta\bigg)V^1 
&\bigg(\frac{1}{\lambda}-\zeta\bigg)V^2 &\dots &\bigg(\frac{1}{\lambda}-\zeta\bigg)V^{N+1}\\
H^1 &- \1& W^{1,2}&\dots&W^{1,N+1}\\
H^2 & W^{2,1}&-\1&\dots & W^{2,N+1}\\
\vdots &\vdots&\vdots&\ddots&\vdots\\
 H^{N+1} &W^{N+1,1}& W^{N+1,2} &\dots &-\1
 \end{BMAT}
 \right]
  \times 
  \left[
   \begin{BMAT}(r)[1pt]{c}{ccc.ccc.c.ccc}
A_{00}\\
\vdots\\
A_{N_M N_M}\\
B^{1}_{00}\\  
 \vdots\\
B^{1}_{N_M N_M}\\
\vdots\\
B^{N+1}_{00}\\  
 \vdots\\
B^{N+1}_{N_M N_M}\\
\end{BMAT}
\right]
   = 
\left[
\begin{BMAT}(r)[1pt]{c}{ccc.ccc.c.ccc}
1\\
\vdots\\
0\\
0\\  
 \vdots\\
0\\
\vdots\\
0\\  
 \vdots\\
0
\end{BMAT}
\right]
\]
\end{widetext}
To solve this system of equation numerically we employ standard linear algebra 
packages (LAPACK). The number of multipoles $N_{M}$ considered to truncate the system 
was chosen so that the relative accuracy on the rate was less than or equal to 
$TOL=10^{-3}$, namely $|k(N_{M}+1) - k(N_{M})|/k(N_{M}) < TOL$. 

%
%
\section{\label{app:B}}

The monopole approximation of the system~\eqref{sistemananoridotto} for a given configuration 
of the nanocatalysts can be obtained by truncating the expansion to $q=n=0$.
The ensuing equations read
\begin{equation}
\label{mononano0}
\left\{
\begin{aligned}
&\frac{A_{00}}{\lambda} + \left( 
                         \frac{1}{\lambda} - \zeta 
                       \right) \sum_{\beta=1}^{N}B_{00}^{\beta}
          V_{00}^{\beta 00} = 1\\
&B_{00}^{\alpha} + \frac{h_\alpha}{1+h_\alpha}
         \bigg( A_{00} H_{00}^{(\alpha 00)}
         +\sum_{\beta\neq \alpha=1}^{N}B_{00}^{\beta} 
                    W_{00}^{(\alpha \beta 00)}\bigg)=0 
\end{aligned}
\right.
\end{equation}
with $\alpha=1,2,\dots, N$. 
Recalling the definitions~\eqref{e:Vdef},~\eqref{e:Hdef} and~\eqref{e:Wdef}, we have 
$V_{00}^{\beta 00} = a/R_0$, $H_{00}^{(\beta 00)}=1$, 
$W_{00}^{(\alpha \beta 00)}=a/L_{\beta\alpha}$, and $\zeta = D_i/D_0$, so that 
Eqs.~\eqref{mononano0} take the following form
\begin{equation}
\label{mononano1}
\left\{
\begin{aligned}
&\frac{A_{00}}{\lambda}  + \frac{a}{R_0}
                       \left( 
                         \frac{1}{\lambda} - \zeta 
                       \right) \sum_{\beta=1}^{N}B_{00}^{\beta} = 1\\
&B_{00}^{\alpha} + \frac{\ds h_\alpha}{\ds 1 + h_\alpha}
         \bigg( A_{00} 
         +\sum_{\beta\neq \alpha=1}^{N}B_{00}^{\beta} 
                   \frac{a}{L_{\alpha\beta}}\bigg)=0
\end{aligned}
\right.
\end{equation}
Since $B^\alpha_{00}= - k_\alpha/k^+_S$,  the overall rate constant of the nanoreactor
can be computed simply as $k = -k^+_S\sum_{\beta=1}^{N+1} B_{00}^{\beta}$
(note that $B_{00}^1=0$ is identically zero as the PS core is modeled 
as a reflecting sphere). 
Moreover, we can average the system~\eqref{mononano1} over the catalyst configurations, 
in the reasonable hypothesis that spatial correlations between the positions of the 
catalysts are negligible. This reduces the many-body average to a two-body problem,
namely 
\begin{widetext}
\begin{eqnarray}
\label{e:averinvLapp}
\left\langle
   \frac{a}{L_{\alpha\beta}}
\right\rangle &=& 
\frac{9\,a}{2[(R_0-a)^3 - (R_S+a)^3]^2} \int_{R_S+a}^{R_0-a} r^2 \, dr
                           \int_{R_S+a}^{R_0-a} \rho^2 \, d\rho
                           \int_0^\pi \frac{\sin \theta}
                                           {\sqrt{r^2 + \rho^2 - 2r\rho\cos\theta}}\,d\theta
                                           \nonumber\\
              &=& \frac{2(1-\ve)^5 - 5 (1-\ve)^2(\gamma+\ve)^3 + 3(\gamma+\ve)^5}
                       {(1-\ve)^6  - 2 (1-\ve)^3(\gamma+\ve)^3 +  (\gamma+\ve)^6}  
                  \left( 
                      \frac{3a}{5R_0} 
                  \right)       
                  := \ve \, C (\ve,\gamma) 
\end{eqnarray}  
\end{widetext}
where $\gamma = R_S/R_0$. We therefore  get from Eqs.~\eqref{mononano1}
\begin{equation}
\label{mononano2}
\left\{
\begin{aligned}
&\frac{A_{00}}{\lambda} - \frac{a}{R_0}
                       \left( 
                         \frac{1}{\lambda} - \zeta 
                       \right) \frac{k}{k^+_S}  = 1 \\
&k - \frac{h}{1 + h}
         \bigg[  N A_{00} \, k^+_S
                 -(N-1) k
                  \left\langle \frac{a}{L_{\alpha\beta}} \right\rangle
         \bigg] = 0 
\end{aligned}
\right.
\end{equation}
where we have taken $h_\alpha = h = k^\ast/k_S^+$ \ $\forall \ \alpha$ as the $N$ 
catalysts are identical. By eliminating $A_{00}$ the solution~\eqref{e:ratef} is easily 
recovered as 
\begin{equation}
\label{e:kfin}
\frac{k}{k_S^-} = \zeta \,\ve  \left( \frac{k}{k_S^+} \right) 
\end{equation}
The diffusion-limited solution~\eqref{e:ratefDL} follows straightforwardly in the limit $h \to \infty$.

%

\begin{mcitethebibliography}{39}
\providecommand*{\natexlab}[1]{#1}
\providecommand*{\mciteSetBstSublistMode}[1]{}
\providecommand*{\mciteSetBstMaxWidthForm}[2]{}
\providecommand*{\mciteBstWouldAddEndPuncttrue}
  {\def\EndOfBibitem{\unskip.}}
\providecommand*{\mciteBstWouldAddEndPunctfalse}
  {\let\EndOfBibitem\relax}
\providecommand*{\mciteSetBstMidEndSepPunct}[3]{}
\providecommand*{\mciteSetBstSublistLabelBeginEnd}[3]{}
\providecommand*{\EndOfBibitem}{}
\mciteSetBstSublistMode{f}
\mciteSetBstMaxWidthForm{subitem}
{(\emph{\alph{mcitesubitemcount}})}
\mciteSetBstSublistLabelBeginEnd{\mcitemaxwidthsubitemform\space}
{\relax}{\relax}

\bibitem[Pushkarev \emph{et~al.}(2012)Pushkarev, Zhu, An, Hervier, and
  Somorjai]{Pushkarev2012}
V.~V. Pushkarev, Z.~Zhu, K.~An, A.~Hervier and G.~A. Somorjai, \emph{Topics in
  Catalysis}, 2012, \textbf{55}, 1257--1275\relax
\mciteBstWouldAddEndPuncttrue
\mciteSetBstMidEndSepPunct{\mcitedefaultmidpunct}
{\mcitedefaultendpunct}{\mcitedefaultseppunct}\relax
\EndOfBibitem
\bibitem[Zhang \emph{et~al.}(2012)Zhang, Cui, Shi, and Deng]{Zhang2012}
Y.~Zhang, X.~Cui, F.~Shi and Y.~Deng, \emph{Chemical Reviews}, 2012,
  \textbf{112}, 2467--2505\relax
\mciteBstWouldAddEndPuncttrue
\mciteSetBstMidEndSepPunct{\mcitedefaultmidpunct}
{\mcitedefaultendpunct}{\mcitedefaultseppunct}\relax
\EndOfBibitem
\bibitem[Haruta(2003)]{Haruta2003}
M.~Haruta, \emph{Chemical Record}, 2003, \textbf{3}, 75--87\relax
\mciteBstWouldAddEndPuncttrue
\mciteSetBstMidEndSepPunct{\mcitedefaultmidpunct}
{\mcitedefaultendpunct}{\mcitedefaultseppunct}\relax
\EndOfBibitem
\bibitem[Hutchings and Haruta(2005)]{Hutchings2005}
G.~J. Hutchings and M.~Haruta, \emph{Applied Catalysis A: General}, 2005,
  \textbf{291}, 2--5\relax
\mciteBstWouldAddEndPuncttrue
\mciteSetBstMidEndSepPunct{\mcitedefaultmidpunct}
{\mcitedefaultendpunct}{\mcitedefaultseppunct}\relax
\EndOfBibitem
\bibitem[Crooks \emph{et~al.}(2001)Crooks, Zhao, Sun, Chechik, and
  Yeung]{Crooks2001}
R.~M. Crooks, M.~Zhao, L.~Sun, V.~Chechik and L.~K. Yeung, \emph{Accounts of
  Chemical Research}, 2001, \textbf{34}, 181--190\relax
\mciteBstWouldAddEndPuncttrue
\mciteSetBstMidEndSepPunct{\mcitedefaultmidpunct}
{\mcitedefaultendpunct}{\mcitedefaultseppunct}\relax
\EndOfBibitem
\bibitem[Noh and Meijboom(2015)]{Noh2015}
J.-H. Noh and R.~Meijboom, \emph{Applied Catalysis A: General}, 2015,
  \textbf{497}, 107--120\relax
\mciteBstWouldAddEndPuncttrue
\mciteSetBstMidEndSepPunct{\mcitedefaultmidpunct}
{\mcitedefaultendpunct}{\mcitedefaultseppunct}\relax
\EndOfBibitem
\bibitem[Ballauff(2007)]{Ballauff2007}
M.~Ballauff, \emph{Progress in Polymer Science}, 2007, \textbf{32},
  1135--1151\relax
\mciteBstWouldAddEndPuncttrue
\mciteSetBstMidEndSepPunct{\mcitedefaultmidpunct}
{\mcitedefaultendpunct}{\mcitedefaultseppunct}\relax
\EndOfBibitem
\bibitem[Lu and Ballauff(2011)]{Lu2011}
Y.~Lu and M.~Ballauff, \emph{Progress in Polymer Science (Oxford)}, 2011,
  \textbf{36}, 767--792\relax
\mciteBstWouldAddEndPuncttrue
\mciteSetBstMidEndSepPunct{\mcitedefaultmidpunct}
{\mcitedefaultendpunct}{\mcitedefaultseppunct}\relax
\EndOfBibitem
\bibitem[Lu \emph{et~al.}(2006)Lu, Mei, Drechsler, and Ballauff]{Lu2006c}
Y.~Lu, Y.~Mei, M.~Drechsler and M.~Ballauff, \emph{Angewandte Chemie -
  International Edition}, 2006, \textbf{45}, 813--816\relax
\mciteBstWouldAddEndPuncttrue
\mciteSetBstMidEndSepPunct{\mcitedefaultmidpunct}
{\mcitedefaultendpunct}{\mcitedefaultseppunct}\relax
\EndOfBibitem
\bibitem[Carregal-Romero \emph{et~al.}(2010)Carregal-Romero, Buurma,
  Perez-Juste, Liz-Marzan, and Herv~es]{marzan1}
S.~Carregal-Romero, N.~J. Buurma, J.~Perez-Juste, L.~M. Liz-Marzan and
  P.~Herv~es, \emph{Chem. Mater.}, 2010, \textbf{22}, 3051--3059\relax
\mciteBstWouldAddEndPuncttrue
\mciteSetBstMidEndSepPunct{\mcitedefaultmidpunct}
{\mcitedefaultendpunct}{\mcitedefaultseppunct}\relax
\EndOfBibitem
\bibitem[Wu \emph{et~al.}(2012)Wu, Dzubiella, Kaiser, Drechsler, Guo, Ballauff,
  and Lu]{Wu2012}
S.~Wu, J.~Dzubiella, J.~Kaiser, M.~Drechsler, X.~Guo, M.~Ballauff and Y.~Lu,
  \emph{Angewandte Chemie - International Edition}, 2012, \textbf{51},
  2229--2233\relax
\mciteBstWouldAddEndPuncttrue
\mciteSetBstMidEndSepPunct{\mcitedefaultmidpunct}
{\mcitedefaultendpunct}{\mcitedefaultseppunct}\relax
\EndOfBibitem
\bibitem[Angioletti-Uberti \emph{et~al.}(2015)Angioletti-Uberti, Lu, Ballauff,
  and Dzubiella]{Angioletti-Uberti2015}
S.~Angioletti-Uberti, Y.~Lu, M.~Ballauff and J.~Dzubiella, \emph{The Journal of
  Physical Chemistry C}, 2015, \textbf{119}, 15723--15730\relax
\mciteBstWouldAddEndPuncttrue
\mciteSetBstMidEndSepPunct{\mcitedefaultmidpunct}
{\mcitedefaultendpunct}{\mcitedefaultseppunct}\relax
\EndOfBibitem
\bibitem[Debye(1942)]{debye42}
P.~Debye, \emph{Trans. Electrochem. Soc.}, 1942, \textbf{92}, 265--272\relax
\mciteBstWouldAddEndPuncttrue
\mciteSetBstMidEndSepPunct{\mcitedefaultmidpunct}
{\mcitedefaultendpunct}{\mcitedefaultseppunct}\relax
\EndOfBibitem
\bibitem[Shi \emph{et~al.}(2014)Shi, Wang, Wang, Ren, Gao, and Wang]{Shi2014}
S.~Shi, Q.~Wang, T.~Wang, S.~Ren, Y.~Gao and N.~Wang, \emph{The journal of
  physical chemistry B}, 2014, \textbf{118}, 7177--86\relax
\mciteBstWouldAddEndPuncttrue
\mciteSetBstMidEndSepPunct{\mcitedefaultmidpunct}
{\mcitedefaultendpunct}{\mcitedefaultseppunct}\relax
\EndOfBibitem
\bibitem[Aditya \emph{et~al.}(2015)Aditya, Pal, and Pal]{Aditya2015}
T.~Aditya, A.~Pal and T.~Pal, \emph{Chemical communications (Cambridge,
  England)}, 2015, \textbf{51}, 9410--31\relax
\mciteBstWouldAddEndPuncttrue
\mciteSetBstMidEndSepPunct{\mcitedefaultmidpunct}
{\mcitedefaultendpunct}{\mcitedefaultseppunct}\relax
\EndOfBibitem
\bibitem[Zhao \emph{et~al.}(2015)Zhao, Feng, Huang, Yang, and Astruc]{Zhao2015}
P.~Zhao, X.~Feng, D.~Huang, G.~Yang and D.~Astruc, \emph{Coordination Chemistry
  Reviews}, 2015, \textbf{287}, 114--136\relax
\mciteBstWouldAddEndPuncttrue
\mciteSetBstMidEndSepPunct{\mcitedefaultmidpunct}
{\mcitedefaultendpunct}{\mcitedefaultseppunct}\relax
\EndOfBibitem
\bibitem[Herves \emph{et~al.}(2012)Herves, P\'{e}rez-Lorenzo, Liz-Marz\'{a}n,
  Dzubiella, Lu, Ballauff, Herv\'{e}s, P\'{e}rez-Lorenzo, Liz-Marz\'{a}n,
  Dzubiella, Lu, and Ballauff]{Herves2012}
P.~Herves, M.~P\'{e}rez-Lorenzo, L.~M. Liz-Marz\'{a}n, J.~Dzubiella, Y.~Lu,
  M.~Ballauff, P.~Herv\'{e}s, M.~P\'{e}rez-Lorenzo, L.~M. Liz-Marz\'{a}n,
  J.~Dzubiella, Y.~Lu and M.~Ballauff, \emph{Chemical Society Reviews}, 2012,
  \textbf{41}, 5577\relax
\mciteBstWouldAddEndPuncttrue
\mciteSetBstMidEndSepPunct{\mcitedefaultmidpunct}
{\mcitedefaultendpunct}{\mcitedefaultseppunct}\relax
\EndOfBibitem
\bibitem[Gu \emph{et~al.}(2014)Gu, Wunder, Lu, Ballauff, Fenger, Rademann,
  Jaquet, and Zaccone]{Gu2014}
S.~Gu, S.~Wunder, Y.~Lu, M.~Ballauff, R.~Fenger, K.~Rademann, B.~Jaquet and
  A.~Zaccone, \emph{The Journal of Physical Chemistry C}, 2014, \textbf{118},
  18618--18625\relax
\mciteBstWouldAddEndPuncttrue
\mciteSetBstMidEndSepPunct{\mcitedefaultmidpunct}
{\mcitedefaultendpunct}{\mcitedefaultseppunct}\relax
\EndOfBibitem
\bibitem[Welsch \emph{et~al.}(2012)Welsch, Becker, Dzubiella, and
  Ballauff]{Welsch2012}
N.~Welsch, A.~L. Becker, J.~Dzubiella and M.~Ballauff, \emph{Soft Matter},
  2012, \textbf{8}, 1428\relax
\mciteBstWouldAddEndPuncttrue
\mciteSetBstMidEndSepPunct{\mcitedefaultmidpunct}
{\mcitedefaultendpunct}{\mcitedefaultseppunct}\relax
\EndOfBibitem
\bibitem[Calef and Deutch(1983)]{Calef:1983aa}
D.~F. Calef and J.~M. Deutch, \emph{Annual Review of Physical Chemistry}, 1983,
  \textbf{34}, 493--524\relax
\mciteBstWouldAddEndPuncttrue
\mciteSetBstMidEndSepPunct{\mcitedefaultmidpunct}
{\mcitedefaultendpunct}{\mcitedefaultseppunct}\relax
\EndOfBibitem
\bibitem[Rice(1985)]{Rice:1985kx}
\emph{Diffusion-limited reactions}, ed. S.~A. Rice, Elsevier, Amsterdam, 1985,
  vol.~25\relax
\mciteBstWouldAddEndPuncttrue
\mciteSetBstMidEndSepPunct{\mcitedefaultmidpunct}
{\mcitedefaultendpunct}{\mcitedefaultseppunct}\relax
\EndOfBibitem
\bibitem[Szabo(1989)]{Szabo:1989fk}
A.~Szabo, \emph{The Journal of Physical Chemistry}, 1989, \textbf{93},
  6929--6939\relax
\mciteBstWouldAddEndPuncttrue
\mciteSetBstMidEndSepPunct{\mcitedefaultmidpunct}
{\mcitedefaultendpunct}{\mcitedefaultseppunct}\relax
\EndOfBibitem
\bibitem[Zhou \emph{et~al.}(2008)Zhou, Rivas, and Minton]{Zhou:2008vf}
H.~X. Zhou, G.~Rivas and A.~P. Minton, \emph{Annual review of biophysics},
  2008, \textbf{37}, 375--397\relax
\mciteBstWouldAddEndPuncttrue
\mciteSetBstMidEndSepPunct{\mcitedefaultmidpunct}
{\mcitedefaultendpunct}{\mcitedefaultseppunct}\relax
\EndOfBibitem
\bibitem[von Smoluchowski(1916)]{Smoluchowski:1916fk}
M.~von Smoluchowski, \emph{Physik Z}, 1916, \textbf{17}, 557--571\relax
\mciteBstWouldAddEndPuncttrue
\mciteSetBstMidEndSepPunct{\mcitedefaultmidpunct}
{\mcitedefaultendpunct}{\mcitedefaultseppunct}\relax
\EndOfBibitem
\bibitem[Collins and Kimball(1949)]{Collins:1949aa}
F.~C. Collins and G.~E. Kimball, \emph{Journal of Colloid Science}, 1949,
  \textbf{4}, 425--437\relax
\mciteBstWouldAddEndPuncttrue
\mciteSetBstMidEndSepPunct{\mcitedefaultmidpunct}
{\mcitedefaultendpunct}{\mcitedefaultseppunct}\relax
\EndOfBibitem
\bibitem[Deutch \emph{et~al.}(1976)Deutch, Felderhof, and
  Saxton]{Deutch:1976aa}
J.~M. Deutch, B.~U. Felderhof and M.~J. Saxton, \emph{The Journal of Chemical
  Physics}, 1976, \textbf{64}, 4559\relax
\mciteBstWouldAddEndPuncttrue
\mciteSetBstMidEndSepPunct{\mcitedefaultmidpunct}
{\mcitedefaultendpunct}{\mcitedefaultseppunct}\relax
\EndOfBibitem
\bibitem[Felderhof and Deutch(1976)]{Felderhof:1976aa}
B.~U. Felderhof and J.~M. Deutch, \emph{The Journal of Chemical Physics}, 1976,
  \textbf{64}, 4551\relax
\mciteBstWouldAddEndPuncttrue
\mciteSetBstMidEndSepPunct{\mcitedefaultmidpunct}
{\mcitedefaultendpunct}{\mcitedefaultseppunct}\relax
\EndOfBibitem
\bibitem[Traytak(1992)]{Traytak:1992}
S.~D. Traytak, \emph{Chemical Physics Letters}, 1992, \textbf{197}, 247 --
  254\relax
\mciteBstWouldAddEndPuncttrue
\mciteSetBstMidEndSepPunct{\mcitedefaultmidpunct}
{\mcitedefaultendpunct}{\mcitedefaultseppunct}\relax
\EndOfBibitem
\bibitem[Traytak(2003)]{Traytak:2003aa}
S.~D. Traytak, \emph{The Journal of Composite Mechanics And Design}, 2003,
  \textbf{9}, 495--521\relax
\mciteBstWouldAddEndPuncttrue
\mciteSetBstMidEndSepPunct{\mcitedefaultmidpunct}
{\mcitedefaultendpunct}{\mcitedefaultseppunct}\relax
\EndOfBibitem
\bibitem[Gordeliy \emph{et~al.}(2009)Gordeliy, Crouch, and
  Mogilevskaya]{Gordeliy:2009aa}
E.~Gordeliy, S.~L. Crouch and S.~G. Mogilevskaya, \emph{International Journal
  for Numerical Methods in Engineering}, 2009, \textbf{77}, 751--775\relax
\mciteBstWouldAddEndPuncttrue
\mciteSetBstMidEndSepPunct{\mcitedefaultmidpunct}
{\mcitedefaultendpunct}{\mcitedefaultseppunct}\relax
\EndOfBibitem
\bibitem[Galanti \emph{et~al.}(2016)Galanti, Fanelli, Traytak, and
  Piazza]{Galanti:2016ab}
M.~Galanti, D.~Fanelli, S.~D. Traytak and F.~Piazza, \emph{Phys. Chem. Chem.
  Phys.}, 2016, \textbf{18}, 15950--15954\relax
\mciteBstWouldAddEndPuncttrue
\mciteSetBstMidEndSepPunct{\mcitedefaultmidpunct}
{\mcitedefaultendpunct}{\mcitedefaultseppunct}\relax
\EndOfBibitem
\bibitem[Cukier(1984)]{Cukier:1984ve}
R.~Cukier, \emph{Macromolecules}, 1984, \textbf{17}, 252--255\relax
\mciteBstWouldAddEndPuncttrue
\mciteSetBstMidEndSepPunct{\mcitedefaultmidpunct}
{\mcitedefaultendpunct}{\mcitedefaultseppunct}\relax
\EndOfBibitem
\bibitem[Ladyzhenskaya and Uralt'seva(1968)]{Ladyzhenskaya:1968dz}
O.~A. Ladyzhenskaya and N.~N. Uralt'seva, \emph{Linear and Quasilinear Elliptic
  Equations}, Academic Press, New York and London, 1968, vol.~46\relax
\mciteBstWouldAddEndPuncttrue
\mciteSetBstMidEndSepPunct{\mcitedefaultmidpunct}
{\mcitedefaultendpunct}{\mcitedefaultseppunct}\relax
\EndOfBibitem
\bibitem[Mei \emph{et~al.}(2007)Mei, Lu, Polzer, Ballauff, and
  Drechsler]{Mei:2007vn}
Y.~Mei, Y.~Lu, F.~Polzer, M.~Ballauff and M.~Drechsler, \emph{Chemistry of
  Materials}, 2007, \textbf{19}, 1062--1069\relax
\mciteBstWouldAddEndPuncttrue
\mciteSetBstMidEndSepPunct{\mcitedefaultmidpunct}
{\mcitedefaultendpunct}{\mcitedefaultseppunct}\relax
\EndOfBibitem
\bibitem[Yigit \emph{et~al.}(2012)Yigit, Welsch, Ballauff, and
  Dzubiella]{Yigit:2012aa}
C.~Yigit, N.~Welsch, M.~Ballauff and J.~Dzubiella, \emph{Langmuir}, 2012,
  \textbf{28}, 14373--14385\relax
\mciteBstWouldAddEndPuncttrue
\mciteSetBstMidEndSepPunct{\mcitedefaultmidpunct}
{\mcitedefaultendpunct}{\mcitedefaultseppunct}\relax
\EndOfBibitem
\bibitem[Palasis and Gehrke(1992)]{Palasis:1992aa}
M.~Palasis and S.~H. Gehrke, \emph{Journal of Controlled Release}, 1992,
  \textbf{18}, 1--11\relax
\mciteBstWouldAddEndPuncttrue
\mciteSetBstMidEndSepPunct{\mcitedefaultmidpunct}
{\mcitedefaultendpunct}{\mcitedefaultseppunct}\relax
\EndOfBibitem
\bibitem[Galanti \emph{et~al.}(2015)Galanti, Fanelli, and
  Piazza]{Galanti:2015aa}
M.~Galanti, D.~Fanelli and F.~Piazza, 2015,  5\relax
\mciteBstWouldAddEndPuncttrue
\mciteSetBstMidEndSepPunct{\mcitedefaultmidpunct}
{\mcitedefaultendpunct}{\mcitedefaultseppunct}\relax
\EndOfBibitem
\bibitem[Arfken \emph{et~al.}(2005)Arfken, Weber, and Harris]{Arfken:2005cr}
G.~Arfken, H.~J. Weber and F.~E. Harris, \emph{Mathematical Methods for
  Physicists, Sixth Edition: A Comprehensive Guide}, Elsevier Academic Press,
  2005\relax
\mciteBstWouldAddEndPuncttrue
\mciteSetBstMidEndSepPunct{\mcitedefaultmidpunct}
{\mcitedefaultendpunct}{\mcitedefaultseppunct}\relax
\EndOfBibitem
\bibitem[Caola(2001)]{Caola:2001pd}
M.~J. Caola, \emph{Journal of Physics A: Mathematical and General}, 2001,
  \textbf{11}, L23--L25\relax
\mciteBstWouldAddEndPuncttrue
\mciteSetBstMidEndSepPunct{\mcitedefaultmidpunct}
{\mcitedefaultendpunct}{\mcitedefaultseppunct}\relax
\EndOfBibitem
\end{mcitethebibliography}
%

\providecommand*{\mcitethebibliography}{\thebibliography}
\csname @ifundefined\endcsname{endmcitethebibliography}
{\let\endmcitethebibliography\endthebibliography}{}

\end{document}